\documentclass[11pt]{article}
\usepackage{epsfig}

\renewcommand{\theequation}{\thesection.\arabic{equation}}
\def\ket#1{\left| #1 \right\rangle}

\begin{document}

\begin{titlepage}
\rightline{\tt hep-th/0603159}
\rightline{\tt MIT-CTP-3727}
\begin{center}
\vskip 2.5cm
{\Large \bf {Comments on Schnabl's analytic solution
for tachyon condensation}}
\vskip 0.2cm
{\Large \bf {in Witten's open string field theory}}
\vskip 1.0cm
{\large {Yuji Okawa}}
\vskip 1.0cm
{\it {Center for Theoretical Physics, Room NE25-4093}}\\
{\it {Massachusetts Institute of Technology}}\\
{\it {Cambridge, MA 02139, USA}}\\
okawa@lns.mit.edu
\vskip 1.8cm
{\bf Abstract}
\end{center}

\noindent
Schnabl recently constructed an analytic solution
for tachyon condensation
in Witten's open string field theory.
The solution consists of two pieces.
Only the first piece is involved
in proving that the solution satisfies the equation of motion
when contracted with any state in the Fock space.
On the other hand,
both pieces contribute
in evaluating the kinetic term
to reproduce the value predicted by Sen's conjecture.
We therefore need to understand
why the second piece is necessary.
We evaluate the cubic term of the string field theory action
for Schnabl's solution
and use it to show that the second piece is necessary
for the equation of motion contracted with the solution itself
to be satisfied.
We also present the solution in various forms
including a pure-gauge configuration
and provide simpler proofs
that it satisfies the equation of motion.

\end{titlepage}

\tableofcontents

\section{Introduction}
\setcounter{equation}{0}

Witten's cubic open string field theory \cite{Witten:1985cc}
has been used to calculate the potential
of the open string tachyon in bosonic string theory,
and convincing evidence has been accumulated
for the existence of a nontrivial critical point in the potential
by an approximation scheme called level truncation
\cite{Kostelecky:1988ta, Kostelecky:1989nt, Kostelecky:1995qk,
Sen:1999nx, Moeller:2000xv, Taylor:2002fy, Gaiotto:2002wy}.
The depth of the potential at the critical point
coincides with the D25-brane tension
with impressive precision \cite{Sen:1999nx, Moeller:2000xv,
Taylor:2002fy, Gaiotto:2002wy},
providing strong support for Sen's conjecture
\cite{Sen:1999mh, Sen:1999xm}.
The equation of motion of Witten's string field theory
\cite{Witten:1985cc} is given by
\begin{equation}
Q_B \Psi + \Psi \ast \Psi = 0 \,,
\end{equation}
where $Q_B$ is the BRST operator,
and the product in the second term is Witten's star product.
The value of the potential at the critical point
predicted by Sen's conjecture is
\begin{equation}
\frac{1}{\alpha'^3 g_T^2}
\left[ \, \frac{1}{2} \, \langle \, \Psi , Q_B \Psi \, \rangle
+ \frac{1}{3} \, \langle \, \Psi , \Psi \ast \Psi \, \rangle \,
\right] 
= {}- \frac{1}{2 \, \pi^2 \alpha'^3 g_T^2} \,,
\end{equation}
where $g_T$ is the on-shell three-tachyon
coupling constant.\footnote{
See appendix A of \cite{Okawa:2002pd}
for a derivation of the relation
between the D25-brane tension and $g_T$ from equations
in Polchinski's book \cite{Polchinski:1998rq}.
}
We only consider translationally invariant
string field configurations in this paper,
and the inner product is defined
to be the standard BPZ inner product
divided by the space-time volume factor.
Combining this with the equation of motion,
the prediction from Sen's conjecture therefore amounts to
\begin{eqnarray}
\langle \, \Psi , Q_B \Psi \, \rangle
&=& {}- \frac{3}{\pi^2} \,,
\label{prediction-kinetic}
\\
\langle \, \Psi , \Psi \ast \Psi \, \rangle \,
&=& \frac{3}{\pi^2} \,.
\label{prediction-cubic}
\end{eqnarray}

Schnabl recently constructed
an analytic solution for the tachyon vacuum
in Witten's string field theory \cite{Schnabl:2005gv}.
The solution $\Psi$ consists of two pieces
and is defined by a limit:
\begin{equation}
\Psi = \lim_{N \to \infty} \left[ \,
\sum_{n=0}^N \frac{d}{dn} \psi_n - \psi_N \, \right]
\equiv \lim_{N \to \infty} \left[ \,
\sum_{n=0}^N \psi'_n - \psi_N \, \right] \,,
\label{Psi}
\end{equation}
where the state $\psi_n$
defined for any real $n$ in the range $n \ge 0$
is made of the wedge state
\cite{Rastelli:2000iu, Rastelli:2001vb, Schnabl:2002gg}
with some operator insertions.\footnote{
Our notion is slightly different from Schnabl's
so that the solution (\ref{Psi}) differs from
that in \cite{Schnabl:2005gv} by an overall sign.
See the beginning of the next section for more details.}
It was shown that the string field $\Psi$ in (\ref{Psi}) satisfies
the equation of motion of Witten's string field theory
contracted with any state $\phi$ in the Fock space:
\begin{equation}
\langle \, \phi , Q_B \Psi \, \rangle
+ \langle \, \phi , \Psi \ast \Psi \, \rangle = 0 \,.
\label{equation-with-phi}
\end{equation}
The kinetic term of the Witten's string field theory action
was then evaluated for the solution $\Psi$,
and the value (\ref{prediction-kinetic}) predicted
by Sen's conjecture was analytically reproduced.

Actually, the $\psi_N$ piece
of the solution $\Psi$ in (\ref{Psi}) does not contribute
to inner products with states in the Fock space. Namely,
\begin{equation}
\lim_{N \to \infty} \langle \, \phi , \psi_N \, \rangle = 0
\end{equation}
for any state $\phi$ in the Fock space.
What was really shown in \cite{Schnabl:2005gv} is
\begin{equation}
\sum_{n=0}^\infty \langle \, \phi , Q_B \psi'_n \, \rangle
+ \sum_{n=0}^\infty \sum_{m=0}^\infty
\langle \, \phi , \psi'_n \ast \psi'_m \, \rangle = 0
\end{equation}
for any state $\phi$ in the Fock space.
In other words, the $\psi_N$ piece is not required
by (\ref{equation-with-phi}).
On the other hand, the $\psi_N$ piece does contribute
in evaluating the kinetic term
of the Witten's string field theory action.
More explicitly, the following quantities were
calculated in \cite{Schnabl:2005gv}:
\begin{eqnarray}
{\cal K}_2
&=& \lim_{N \to \infty} \sum_{n=0}^N \sum_{m=0}^N \,
\langle \, \psi'_n , Q_B \psi'_m \, \rangle
= \frac{1}{2} - \frac{1}{\pi^2} \,,
\\
{\cal K}_1
&=& \lim_{N \to \infty} \sum_{m=0}^N \,
\langle \, \psi_N , Q_B \psi'_m \, \rangle
= \frac{1}{2} + \frac{2}{\pi^2} \,,
\\
{\cal K}_0
&=& \lim_{N \to \infty}
\langle \, \psi_N , Q_B \psi_N \, \rangle
= \frac{1}{2} + \frac{2}{\pi^2} \,.
\end{eqnarray}
The inner product $\langle \, \Psi , Q_B \Psi \, \rangle$
for the solution $\Psi$ in (\ref{Psi}) is then
\begin{equation}
\langle \, \Psi , Q_B \Psi \, \rangle
= {\cal K}_2 - 2 \, {\cal K}_1 + {\cal K}_0
= {}- \frac{3}{\pi^2} \,.
\label{prediction-kinetic-reproduced}
\end{equation}
In order to reproduce the value (\ref{prediction-kinetic})
predicted by Sen's conjecture,
the $\psi_N$ piece is really necessary.
In particular, the coefficient in front of $\psi_N$ in (\ref{Psi})
must be $-1$.

In fact, Schnabl first hypothesized the solution
in the following form:
\begin{equation}
\Psi = {}- \sum_{n=0}^\infty \frac{B_n}{n!}
\frac{d^n}{dm^n} \psi_m \biggr|_{m=0} \,,
\label{Psi-Bernoulli}
\end{equation}
where $B_n$'s are the Bernoulli numbers.
It was then brought to the form (\ref{Psi})
using the Euler-Maclaurin formula.
In this sense, it is natural to add the $\psi_N$ piece
with the coefficient $-1$.
However, this is not the requirement that
the equation of motion be satisfied.
If the $\psi_N$ piece is irrelevant
to the equation of motion,
solutions of the form
\begin{equation}
\lim_{N \to \infty} \left[ \,
\sum_{n=0}^N \psi'_n - \alpha \, \psi_N \, \right]
\end{equation}
with any real $\alpha$ will be equally valid.
The evaluation of the kinetic term is well defined
for any $\alpha$, and it yields a value
inconsistent with Sen's conjecture unless $\alpha =1$.
Let us state the puzzle as follows.
\begin{quote}
{\it 1. Why do we need the $\psi_N$ piece?
What determines the coefficient in front of $\psi_N$?}
\end{quote}

There is another question, which turns out to be related
to the above puzzle.
In evaluating the Witten's string field theory action
in \cite{Schnabl:2005gv},
it was implicitly assumed that the equation of motion is
satisfied when it is contracted with the solution itself:
\begin{equation}
\langle \, \Psi , Q_B \Psi \, \rangle
+ \langle \, \Psi , \Psi \ast \Psi \, \rangle = 0 \,.
\label{equation-with-Psi}
\end{equation}
However, this is notoriously subtle in string field theory
because the solution is usually outside the Fock space.
For example, the formal exact solution based on the identity state
\cite{Takahashi:2002ez, Kishimoto:2002xi}
satisfies the equation of motion
when contracted with any state in the Fock space,
but the calculations of the inner products
in (\ref{equation-with-Psi})
are not even well defined.
Another example is the twisted butterfly state
\cite{Gaiotto:2001ji, Schnabl:2002ff, Gaiotto:2002kf}
in vacuum string field theory
\cite{Rastelli:2000hv, Rastelli:2001jb, Rastelli:2001uv}.
It solves the equation of motion
when contracted with any state in the Fock space,
but it does not satisfy the equation
when contracted with the solution itself \cite{Okawa:2003cm},
which indicates that
the assumption of the matter-ghost factorization
in vacuum string field theory
needs to be reconsidered \cite{Drukker:2005hr}.
While a systematic approach to accomplish
the compatibility of (\ref{equation-with-phi})
and (\ref{equation-with-Psi}) has been developed
in \cite{Okawa:2003zc, Yang:2004xz, Drukker:2005hr},
it relies on a series expansion
and the compatibility is only approximate.
It is therefore crucially important
whether or not Schnabl's solution
satisfies (\ref{equation-with-Psi}).
Our question is as follows.
\begin{quote}
{\it 2. Does the solution satisfy the equation of motion
even when it is contracted with the solution itself?}
\end{quote}

In order to address this question, it is necessary
to evaluate the cubic term
of the string field theory action for Schnabl's solution
to see if the value (\ref{prediction-cubic}) is reproduced.
We evaluate the following quantities in this paper:
\begin{eqnarray}
{\cal V}_3
&=& \lim_{N \to \infty} \sum_{n=0}^N \sum_{m=0}^N \sum_{k=0}^N \,
\langle \, \psi'_n , \psi'_m \ast \psi'_k \, \rangle
= \frac{3}{\pi^2} - \frac{3 \sqrt{3}}{2 \pi} \,,
\\
{\cal V}_2
&=& \lim_{N \to \infty} \sum_{m=0}^N \sum_{k=0}^N \,
\langle \, \psi_N , \psi'_m \ast \psi'_k \, \rangle
= {}- \frac{3 \sqrt{3}}{2 \pi} \,,
\\
{\cal V}_1
&=& \lim_{N \to \infty} \sum_{n=0}^N \,
\langle \, \psi'_n , \psi_N \ast \psi_N \, \rangle
= {}- \frac{3 \sqrt{3}}{2 \pi} \,,
\\
{\cal V}_0
&=& \lim_{N \to \infty}
\langle \, \psi_N , \psi_N \ast \psi_N \, \rangle
= {}- \frac{3 \sqrt{3}}{2 \pi} \,.
\end{eqnarray}
The inner product $\langle \, \Psi , \Psi \ast \Psi \, \rangle$
for the solution $\Psi$ in (\ref{Psi}) is then given by
\begin{equation}
\langle \, \Psi , \Psi \ast \Psi \, \rangle \,
= {\cal V}_3 -3 \, {\cal V}_2 +3 \, {\cal V}_1 - {\cal V}_0
= \frac{3}{\pi^2} \,.
\end{equation}
The equation of motion (\ref{equation-with-Psi})
is {\it not} satisfied without the $\psi_N$ piece
because ${\cal K}_2 + {\cal V}_3 \ne 0$,
but it {\it is} nontrivially
satisfied when the $\psi_N$ piece is included
with the coefficient $-1$.
This is the main result of the paper.
We believe that this dispels the questions raised earlier
and provides nontrivial evidence for Schnabl's solution.

Schnabl's solution will definitely play an important role
in developing vacuum string field theory further
or in constructing different analytic solutions
of Witten's string field theory.
It is not clear at this point, however,
what aspects of the solution will be crucial
for future development.
We therefore present the solution
in various forms
including a pure-gauge configuration
and provide simpler proofs
that it satisfies the equation of motion.
We hope that this helps understand the solution better.

The organization of the paper is as follows.
In section \ref{psi_n-section},
we present the string states
$\psi_n$ and $\psi'_n$
in the conformal field theory (CFT) formulation
of string field theory \cite{LeClair:1988sp, LeClair:1988sj}.
We show that Schnabl's solution
satisfies the equation of motion
first in the CFT formulation in subsection \ref{CFT-subsection},
and then we present a purely algebraic proof
in subsection \ref{algebraic-subsection}.
We also express Schnabl's solution
in the half-string picture
in subsection \ref{half-string-subsection}
and as a pure-gauge configuration
in subsection \ref{pure-gauge-subsection}.
The evaluation of the kinetic term
of the string field theory action
for Schnabl's solution in \cite{Schnabl:2005gv}
is reproduced in a different way
in section \ref{Kinetic-term-section},
and the cubic term
of the string field theory action
is evaluated for Schnabl's solution
in section \ref{Cubic-term-section}.
Section \ref{Conclusions} is devoted to conclusions.

\section{Wedge state with operator insertions}
\label{psi_n-section}
\setcounter{equation}{0}

Let us first explain the difference
between Schnabl's notation and ours.
Schnabl's {\it left} is our {\it right}
and Schnabl's {\it right} is our {\it left}.
When an operator ${\cal O}$ is defined
by an integral along the unit circle,
\begin{equation}
{\cal O} = \oint \frac{d \xi}{2 \pi i} \, \varphi (\xi) \,,
\end{equation}
where $\varphi (\xi)$ can be a field
or a field multiplied by a function of $\xi$,
we define ${\cal O}^R$ and ${\cal O}^L$ by
\begin{equation}
{\cal O}^R
= \int_{C_R} \frac{d \xi}{2 \pi i} \, \varphi (\xi) \,, \qquad
{\cal O}^L
= \int_{C_L} \frac{d \xi}{2 \pi i} \, \varphi (\xi) \,,
\end{equation}
where the contour $C_R$ runs
along the right half of the unit circle from $-i$ to $i$
counterclockwise
and the contour $C_L$ runs
along the left half of the unit circle from $i$ to $-i$
counterclockwise.
Schnabl's ${\cal O}^L$ is our ${\cal O}^R$,
and Schnabl's ${\cal O}^R$ is our ${\cal O}^L$.
In this paper, we use the following operators of these forms:
\begin{eqnarray}
&& B_1^R = \int_{C_R} \frac{d \xi}{2 \pi i} \,
(\xi^2+1) \, b (\xi) \,, \qquad
B_1^L = \int_{C_L} \frac{d \xi}{2 \pi i} \,
(\xi^2+1) \, b (\xi) \,,
\label{B_1^R-B_1^L}
\\
&& K_1^R = \int_{C_R} \frac{d \xi}{2 \pi i} \,
(\xi^2+1) \, T (\xi) \,, \qquad
K_1^L = \int_{C_L} \frac{d \xi}{2 \pi i} \,
(\xi^2+1) \, T (\xi) \,,
\label{K_1^R-K_1^L}
\end{eqnarray}
where $b(\xi)$ is the $b$ ghost
and $T(\xi)$ is the energy-momentum tensor.

This difference in the definition of $left$ and $right$
also affects the definition of the star product.
Schnabl's $A \ast B$ is our $(-1)^{A B} B \ast A \,$.\footnote{
In our definition,
the {\it left} string field $A$ in $A \ast B$
is mapped to the {\it left} of the string field $B$
in the complex plane for the glued surface
in the CFT formulation,
and the operators associated with $A$
are located to the {\it left}
of those associated with $B$
in correlation functions
used in the CFT formulation.
See (\ref{star-product-example}), for example.
Schnabl adopted the definition used
in a series of papers
by Rastelli, Sen and Zwiebach 
such as \cite{Rastelli:2001vb}
or in a review \cite{Taylor:2003gn}.}
Here and in what follows a string field in the exponent
of $-1$ denotes its Grassmann property:
it is $0$ mod $2$ for a Grassmann-even string field
and is $1$ mod $2$ for a Grassmann-odd string field.
Schnabl solved the equation
$Q_B \Psi + \Psi \ast \Psi = 0 \,$,
which corresponds to
$Q_B \Psi - \Psi \ast \Psi = 0$
in our notation.
We solve
$Q_B \Psi + \Psi \ast \Psi = 0$
in our notation
so that the solution in this paper should differ from
that in \cite{Schnabl:2005gv}
by an overall minus sign.

The string state $\psi_n$
introduced by Schnabl in \cite{Schnabl:2005gv}
takes the form of the wedge state with some operator insertions.
Let us review the definition of the wedge state
\cite{Rastelli:2000iu, Rastelli:2001vb, Schnabl:2002gg}.
The wedge state $\ket{n}$
for any real $n$ in the range $n > 1$ can be defined by
its inner products with states in the Fock space.
For any state $\phi$ in the Fock space,
the inner product $\langle \, \phi , n \, \rangle$ is given by
\begin{equation}
\langle \, \phi , n \, \rangle
= \langle \, f_n \circ \phi (0) \, \rangle_{\rm UHP} \,,
\label{wedge-definition}
\end{equation}
where $\phi (0)$ is the operator corresponding to the state $\phi$
in the state-operator mapping.
We use the notation $f \circ {\cal O} (\xi)$
for the operator mapped from ${\cal O} (\xi)$
by a conformal transformation $f(\xi)$.
When the operator ${\cal O}$ is primary with dimension $h$,
$f \circ {\cal O} (\xi)$ is as follows:
\begin{equation}
f \circ {\cal O} (\xi)
= \left( \frac{d f(\xi)}{d \xi} \right)^h
{\cal O} ( f(\xi) ) \,.
\end{equation}
The conformal transformation $f_n (\xi)$
in (\ref{wedge-definition}) is given by
\begin{equation}
f_n (\xi)
= \frac{n}{2} \tan \left( \frac{2}{n} \arctan \xi \right) \,.
\end{equation}
The correlation function is evaluated on the upper-half plane
as indicated by the subscript UHP.
As in the definition of the inner product,
we divide correlation functions
by the overall space-time volume factor.
Our normalization of correlation functions is given by
\begin{equation}
\langle \, c (w_1) \, c (w_2) \, c (w_3) \, \rangle_{\rm UHP}
= (w_1-w_2)(w_1-w_3)(w_2-w_3) \,,
\label{normalization}
\end{equation}
where $c (w)$ is the $c$ ghost.
We use the doubling trick throughout the paper.
The normalization of the state-operator mapping is fixed by
the condition that the $SL(2,R)$-invariant vacuum $\ket{0}$
corresponds to the identity operator.
The normalization of the inner product is also fixed
by this together with (\ref{normalization}).

The inner product (\ref{wedge-definition}) can also be written
in terms of a correlation function on a semi-infinite cylinder.
We denote the semi-infinite cylinder
obtained from the upper-half plane of $z$ with
the identification $z+\ell \simeq z$ by $C_\ell$.
The inner product $\langle \, \phi , n \, \rangle$
in terms of a correlation function
on $C_{\frac{\pi n}{2}}$ is given by
\begin{equation}
\langle \, \phi , n \, \rangle
= \langle \, f_\infty \circ \phi (0) \,
\rangle_{C_{\frac{\pi n}{2}}} \,,
\label{wedge-cylinder}
\end{equation}
where
\begin{equation}
f_\infty (\xi) = \arctan \xi \,.
\end{equation}
We in fact mostly use
this expression of $\langle \, \phi , n \, \rangle$ in this paper.
The expression (\ref{wedge-definition}) can be derived from this
in the following way.
First map $C_{\frac{\pi n}{2}}$ to $C_{\pi}$
by the dilatation $z' = \frac{2}{n} \, z$, where $z = \arctan \xi$
is the coordinate on $C_{\frac{\pi n}{2}}$.
Then the conformal transformation $z'' = \tan z'$ maps
$C_\pi$ to the upper-half plane.
We can further perform the dilatation $w = \frac{n}{2} \, z''$
which maps the upper-half plane to itself
so that the combined transformation $w = f_n (\xi)$
has a limit as $n \to \infty$.
After these three transformations,
the inner product $\langle \, \phi , n \, \rangle$
is given by (\ref{wedge-definition}) in the $w$ coordinate.

The $z$ coordinate given by $z = \arctan \xi$ is very useful
in dealing with the star product
\cite{Rastelli:2000iu, Rastelli:2001vb, Schnabl:2005gv}.
The upper half of the unit disk is mapped
to a semi-infinite strip with a width of $\pi/2$
by the conformal transformation $z = \arctan \xi$,
and the left and right halves of the open string
are mapped to semi-infinite lines parallel to the imaginary axis.
The right half of one string
and the left half of the other string
are glued together in Witten's star product
so that star products of states in the Fock space
can be obtained simply by translation in the $z$ coordinate.
Inner products of states in the Fock space
are also simple in the $z$ coordinate,
and they are obtained by first taking the star product
of the two states and then by gluing together
the left and right halves of the resulting string state.
The vacuum state $\ket{0}$ corresponds
to a semi-infinite strip with a width of $\pi/2$,
and no operators are inserted.
The state $\phi$ in the Fock space corresponds
to a semi-infinite strip with a width of $\pi/2$
which has an insertion of $f_\infty \circ \phi (0)$ at the origin.
If we subtract the piece coming from the state $\phi$
in the expression of $\langle \, \phi , n \, \rangle$
in (\ref{wedge-cylinder}),
the remaining surface is a semi-infinite strip
with a width of $\pi (n-1)/2$,
and there are no operator insertions.
When $n$ is an integer, the wedge state $\ket{n}$ therefore
coincides with a product of vacuum states:
\begin{equation}
\ket{n} = \underbrace{\ket{0} \ast \ket{0} \ast \ldots
\ast \ket{0}}_{n-1} \,.
\end{equation}

\begin{figure}[htb]
\centerline{\epsfxsize=3in\epsfbox{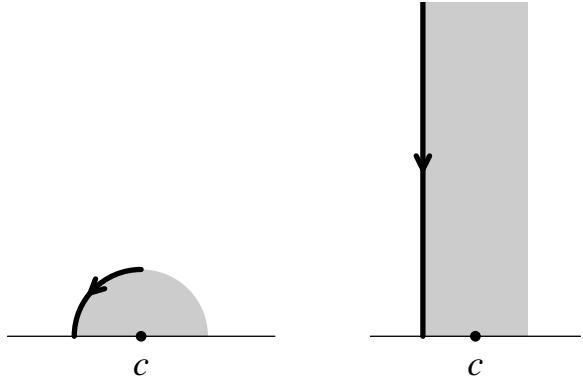}}
\caption{\small
Representations of the state $B_1^L c_1 \ket{0}$.
The left figure is the representation
in the local coordinate $\xi$.
The state $c_1 \ket{0}$ corresponds to the operator $c(0)$,
and the thick line represents
the contour of the integral for $B_1^L$
defined in (\ref{B_1^R-B_1^L}).
The contour runs from $i$ to $-1$
along the unit circle counterclockwise
before using the doubling trick.
The right figure is the representation
after the conformal transformation
$z = f_\infty (\xi) = \arctan \xi$.
The $c$ ghost at the origin remains the same
because $f_\infty \circ c (0) = c(0)$.
The operator $B_1^L$ is mapped to $B$
defined in (\ref{B-definition}),
and the contour of the integral
represented by the thick line
is a semi-infinite line
before using the doubling trick.}
\label{figure1}
\end{figure}

\begin{figure}[htb]
\centerline{\epsfxsize=3in\epsfbox{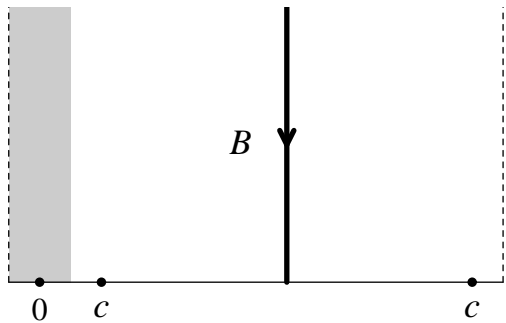}}
\caption{\small
A representation of the inner product
$\langle \, \phi , \psi_n  \, \rangle$.
(The overall factor $2/\pi$ is ignored.)
The two dashed lines are identified.
The shaded region corresponds to the state $\phi$,
and the operator $f_\infty \circ \phi (0)$
is inserted at the origin.
The string state $\psi_n$ is represented
by the rest of the region.
It is made of the wedge state $\ket{n+2}$
with two $c$-ghost insertions
and one insertion of $B$ defined in (\ref{B-definition}).
The contour of the integral
before using the doubling trick
is represented by the thick line.}
\label{figure2}
\end{figure}

The string field $\psi_n$
defined for any real $n$ in the range $n \ge 0$
is made of the wedge state $\ket{n+2}$
with some operator insertions.
It can also be defined
by its inner products with states in the Fock space,
and those inner products can be expressed
by correlation functions
on the semi-infinite cylinder $C_{\pi (n+2)/2}$.
Let us start with the following expression of $\psi_n$
for $n > 1$ given in footnote 16
of \cite{Schnabl:2005gv}:
\begin{equation}
\psi_n = \frac{1}{\pi} \,
c_1 \ket{0} \ast ( B_1^L - B_1^R ) \ket{n} \ast c_1 \ket{0} \,.
\label{Schnabl's-psi_n}
\end{equation}
This takes the same form
both in Schnabl's notation and in ours.\footnote{
Schnabl's $B_1^L - B_1^R$ corresponds to our $B_1^R - B_1^L$,
but the relative minus sign is compensated
when the ordering of the star product is reversed.
}
The mode expansion of the $c$ ghost is given by
\begin{equation}
c_n = \oint \frac{d \xi}{2 \pi i} \, \xi^{n-2} \, c(\xi) \,,
\end{equation}
where the contour encircles the origin counterclockwise.
The state $c_1 \ket{0}$ corresponds to the operator $c(0)$
in the state-operator mapping.
Since $f_\infty \circ c (0) = c(0)$,
the state $c_1 \ket{0}$ is represented
by a semi-infinite strip with a width of $\pi/2$
which has an insertion of $c(0)$
at the midpoint of the finite edge of the strip.
The operator $B_1^L$ defined in (\ref{B_1^R-B_1^L})
is mapped to
\begin{equation}
B = \int_{i \infty}^{-i \infty} \frac{dz}{2 \pi i} \, b (z)
\label{B-definition}
\end{equation}
by the conformal transformation
$z = f_\infty (\xi) = \arctan \xi$.
Note that the function $\xi^2+1$ in (\ref{B_1^R-B_1^L})
is precisely canceled by the conformal factor.
The operator $B$ is therefore invariant under translation
$z \to z + a$ for any real $a$.
The state $B_1^L \ket{\phi}$
for any state $\ket{\phi}$ in the Fock space is represented
by a semi-infinite strip with a width of $\pi/2$,
where the operators $B \, f_\infty \circ \phi (0)$ are inserted.
The contour of the integral in $B$
must be located to the left of $f_\infty \circ \phi (0)$.
See figure \ref{figure1}.
Similarly, the state $B_1^R \ket{\phi}$
for any state $\ket{\phi}$ in the Fock space is represented
by a semi-infinite strip with a width of $\pi/2$,
where the operators
${}-(-1)^{\phi} f_\infty \circ \phi (0) \, B$
are inserted.
Note that the first minus sign comes from
reversing the contour of the integral in $B_1^R$.
The contour of the integral in $B$
must be located to the right of $f_\infty \circ \phi (0)$
in this case.
It is easy to see from this representation that
$B_1^L \ket{0} = - B_1^R \ket{0}$
because no operators are inserted for the vacuum state $\ket{0}$
so that the contour of the integral in $B$ can be
freely deformed inside the semi-infinite strip.
This generalizes to the wedge state,
and the relation $B_1^L \ket{n} = - B_1^R \ket{n}$
holds for any $n$.
Therefore, the factor $( B_1^L - B_1^R ) \ket{n}$
in (\ref{Schnabl's-psi_n})
is equal to $2 \, B_1^L \ket{n}$.
We are now ready to express the state $\psi_n$
in the CFT formulation.
For any state $\phi$ in the Fock space,
the inner product $\langle \, \phi , \psi_n  \, \rangle$
is given by
\begin{equation}
\langle \, \phi , \psi_n \, \rangle
= \frac{2}{\pi} \, \left\langle \,
f_\infty \circ \phi (0) \, c \left( \frac{\pi}{2} \right) \,
B \, c \left( \frac{\pi (n+1)}{2} \right) \,
\right\rangle_{C_{\frac{\pi (n+2)}{2}}} \,.
\label{psi_n-CFT}
\end{equation}
The operator $B$ is defined in (\ref{B-definition}),
and if $B$ is located between two operators,
the contour of the integral must run between the two operators.
In the case of (\ref{psi_n-CFT}), the contour must run
between the two points $\pi/2$ and $\pi (n+1)/2$.
See figure \ref{figure2}.
The semi-infinite cylinder $C_{\pi (n+2)/2}$
can be represented by the region
$-\pi/4 \le {\rm Re} \, z \le \pi (2 \, n + 3)/4$
of the upper-half complex $z$ plane
with the semi-infinite lines at both ends identified,
where ${\rm Re} \, z$ is the real part of $z$.
The region $-\pi/4 \le {\rm Re} \, z \le \pi/4$
with the insertion of $f_\infty \circ \phi (0)$
corresponds to the state $\phi$,
the region $\pi/4 \le {\rm Re} \, z \le 3\, \pi/4$
with a $c$-ghost insertion
corresponds to $c_1 \ket{0}$,
the region $3\, \pi/4 \le {\rm Re} \, z \le \pi (2 \, n + 1)/4$
with the insertion of $B$
corresponds to $B_1^L \ket{n}$,
and the region
$\pi (2 \, n + 1)/4 \le {\rm Re} \, z \le \pi (2 \, n + 3)/4$
with a $c$-ghost insertion
corresponds to $c_1 \ket{0}$.
Since the contour of the integral in $B$
can be deformed as long as it passes
between $\pi/2$ and $\pi (n+1)/2$,
the string field $\psi_n$ with $n > 1$ can also be written as
\begin{equation}
\psi_n = \frac{2}{\pi} \,
c_1 \ket{0} \ast \ket{n} \ast B_1^L c_1 \ket{0} \,.
\label{psi_n-algebraic}
\end{equation}
The wedge state $\ket{n}$ becomes singular when $n < 1$,
but the expression (\ref{psi_n-CFT}) is well defined
in the range $n > 0$.
We define $\psi_n$ for $n > 0$ by (\ref{psi_n-CFT}),
and our definition coincides with that in \cite{Schnabl:2005gv}.
The string field $\psi_1$ is given by
\begin{equation}
\psi_1 = \frac{2}{\pi} \,
c_1 \ket{0} \ast B_1^L c_1 \ket{0} \,.
\end{equation}
The string field $\psi_0$ can be defined by a limit:
\begin{equation}
\psi_0 \equiv \lim_{n \to 0} \psi_n \,.
\end{equation}
Let us calculate $\psi_0$ explicitly.
The anticommutation relation of $B$ and $c(z)$ is given by
\begin{equation}
\{ \, B \,,\, c(z) \, \} = 1 \,.
\label{Bc+cB=1}
\end{equation}
Since
\begin{equation}
\lim_{\epsilon \to 0} c (z) \, B \, c (z+\epsilon)
= c (z) - \lim_{\epsilon \to 0} c (z) \, c (z+\epsilon) \, B
= c (z) \,,
\end{equation}
the inner product $\langle \, \phi , \psi_n \, \rangle$
in the limit $n \to 0$ is given by
\begin{equation}
\lim_{n \to 0} \langle \, \phi , \psi_n \, \rangle
= \frac{2}{\pi} \, \left\langle \,
f_\infty \circ \phi (0) \, c \left( \frac{\pi}{2} \right) \,
\right\rangle_{C_{\pi}} \,.
\end{equation}
We therefore obtain
\begin{equation}
\psi_0 = \frac{2}{\pi} \, c_1 \ket{0} \,.
\end{equation}

\begin{figure}[htb]
\centerline{\epsfxsize=3in\epsfbox{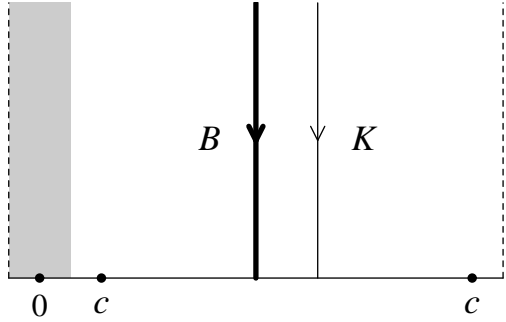}}
\caption{\small
A representation of the inner product
$\langle \, \phi , \psi'_n  \, \rangle$.
The two dashed lines are identified.
The shaded region corresponds to the state $\phi$,
and the operator $f_\infty \circ \phi (0)$
is inserted at the origin.
The string state $\psi'_n$ is represented
by the rest of the region.
It is made of $\psi_n$
with an additional insertion
of $K$ (\ref{K-definition}) multiplied by $\pi/2$.
The contour of the integral for $K$
before using the doubling trick
is represented by the thin line.}
\label{figure3}
\end{figure}

Let us next consider the derivative of $\psi_n$
with respect to $n$.
Since the wedge state
can be written as \cite{Schnabl:2005gv, Schnabl:2002gg}
\begin{equation}
\ket{n} = e^{\frac{\pi (n-2)}{2} K_1^L} \ket{0} \,,
\end{equation}
the derivative of $\ket{n}$
with respect of $n$ is given by
\begin{equation}
\frac{d}{dn} \ket{n}
= \frac{\pi}{2} \, K_1^L \ket{n} \,.
\label{K_1^L-insertion}
\end{equation}
The derivative of $\psi_n$ is then given by
\begin{equation}
\psi'_n \equiv \frac{d}{dn} \psi_n
= c_1 \ket{0} \ast K_1^L \ket{n} \ast B_1^L c_1 \ket{0}
\end{equation}
for $n > 1$.
Just as the operator $e^{- \tau L_0}$ creates
a piece of the open string world-sheet
with a length of $\tau$ in the strip coordinate,
the operator $e^{\, \ell \, K_1^L}$ creates
a semi-infinite strip with a width of $\ell$.
The correspondence between the derivative with respect to $n$
and an insertion of $K_1^L$ multiplied by $\pi/2$
is therefore valid in the whole range $n > 0$.
The operator $K_1^L$ defined in (\ref{K_1^R-K_1^L})
is mapped to
\begin{equation}
K = \int_{i \infty}^{-i \infty} \frac{dz}{2 \pi i} \, T (z)
\label{K-definition}
\end{equation}
by the conformal transformation
$z = f_\infty (\xi) = \arctan \xi$.
Note that $K$ and $B$ commute.
The inner product of $\psi'_n$
with any state $\phi$ in the Fock space is given by
\begin{equation}
\langle \, \phi , \psi'_n \, \rangle
= \left\langle \, f_\infty \circ \phi (0) \,
c \left( \frac{\pi}{2} \right) \,
B \, K \, c \left( \frac{\pi (n+1)}{2} \right) \,
\right\rangle_{C_{\frac{\pi (n+2)}{2}}}
\label{psi'_n-CFT}
\end{equation}
for $n > 0$.
See figure \ref{figure3}.
As in the case of $B$,
if $K$ is located between two operators,
the contour of the integral must run between the two operators.
In the case of (\ref{psi'_n-CFT}), the contour must pass
between the two points $\pi/2$ and $\pi (n+1)/2$.
Recall that $K$ and $B$ commute.
The state $K_1^L \ket{\phi}$
for any state $\ket{\phi}$ in the Fock space is represented
by a semi-infinite strip with a width of $\pi/2$,
where the operators $K \, f_\infty \circ \phi (0)$ are inserted.
The contour of the integral in $K$
must be located to the left of $f_\infty \circ \phi (0)$.
Similarly, the state $K_1^R \ket{\phi}$
for any state $\ket{\phi}$ in the Fock space is represented
by a semi-infinite strip with a width of $\pi/2$,
where the operators ${}- f_\infty \circ \phi (0) \, K$
are inserted.
Note that the minus sign comes from
reversing the contour of the integral in $K_1^R$.
The contour of the integral in $K$
must be located to the right of $f_\infty \circ \phi (0)$
in this case.

The string field $\psi'_n$ can also be written as
\begin{equation}
\psi'_n = c_1 \ket{0} \ast \ket{n} \ast B_1^L K_1^L c_1 \ket{0}
\end{equation}
for $n > 1$, and $\psi'_1$ is
\begin{equation}
\psi'_1 = c_1 \ket{0} \ast B_1^L K_1^L c_1 \ket{0} \,.
\end{equation}
As in the case of $\psi_0$,
the string field $\psi'_0$ can be defined by a limit:
\begin{equation}
\psi'_0 = \lim_{n \to 0} \psi'_n \,.
\end{equation}
Since
\begin{equation}
\lim_{\epsilon \to 0} \, c (z) \, K \, c(z+\epsilon)
= \lim_{\epsilon \to 0} \, \left[ \,
c (z) \, \partial c(z+\epsilon)
+ c (z) \, c(z+\epsilon) \, K \, \right]
= c \partial c (z) \,,
\label{c-K-c}
\end{equation}
the product of operators
$c (z) \, B \, K \, c(z+\epsilon)$
in the limit $\epsilon \to 0$ can be written as
\begin{equation}
\lim_{\epsilon \to 0} \,
c (z) \, B \, K \, c(z+\epsilon)
= \lim_{\epsilon \to 0} \,
K \, c(z+\epsilon) - \lim_{\epsilon \to 0} \,
B \, c (z) \, K \, c(z+\epsilon)
= K \, c (z) - B \, c \partial c (z)
\end{equation}
or as
\begin{equation}
\lim_{\epsilon \to 0} \,
c (z) \, B \, K \, c(z+\epsilon)
= c (z) \, K - \lim_{\epsilon \to 0} \,
c (z) \, K \, c(z+\epsilon) \, B
= c (z) \, K - c \partial c (z) \, B \,.
\label{c-B-K-c-2}
\end{equation}
Using these formulas,
the inner product (\ref{psi'_n-CFT})
in the limit $n \to 0$ is given by
\begin{equation}
\langle \, \phi , \psi'_0 \, \rangle
= \lim_{n \to 0} \langle \, \phi , \psi'_n \, \rangle
= \left\langle \, f_\infty \circ \phi (0) \,
K \, c \left( \frac{\pi}{2} \right) \,
\right\rangle_{C_{\pi}}
- \left\langle \, f_\infty \circ \phi (0) \,
B \, c \partial c \left( \frac{\pi}{2} \right) \,
\right\rangle_{C_{\pi}}
\label{psi'_0-inner-product-1}
\end{equation}
or by
\begin{equation}
\langle \, \phi , \psi'_0 \, \rangle
= \lim_{n \to 0} \langle \, \phi , \psi'_n \, \rangle
= \left\langle \, f_\infty \circ \phi (0) \,
c \left( \frac{\pi}{2} \right) \, K \, 
\right\rangle_{C_{\pi}}
- \left\langle \, f_\infty \circ \phi (0) \,
c \partial c \left( \frac{\pi}{2} \right) \, B \,
\right\rangle_{C_{\pi}} \,.
\label{psi'_0-inner-product-2}
\end{equation}
The string field $\psi'_0$ is therefore given by
\begin{equation}
\psi'_0 = K_1^L c_1 \ket{0} + B_1^L c_0 c_1 \ket{0}
\end{equation}
or by
\begin{equation}
\psi'_0 = - K_1^R c_1 \ket{0} - B_1^R c_0 c_1 \ket{0} \,.
\end{equation}
Note that the operator $c \partial c (0)$ corresponds
to the state ${}- c_0 c_1 \ket{0}$.

\section{Equation of motion}
\label{Equation-of-motion}
\setcounter{equation}{0}

The BRST transformation of the string field $\psi'_n$
with integer $n$ is given by
\begin{eqnarray}
Q_B \psi'_0 &=& 0 \,,
\label{psi'_0-relation}
\\
Q_B \psi'_{n+1}
&=& - \sum_{m=0}^{n} \psi'_{m} \ast \psi'_{n-m}
\label{psi'_n-relation}
\end{eqnarray}
for $n \ge 0$.
Because of this remarkable property,
the string field $\Psi_\lambda$ given by
\begin{equation}
\Psi_\lambda = \sum_{n=0}^\infty \lambda^{n+1} \, \psi'_n
\label{lambda}
\end{equation}
for any real $\lambda$ formally satisfies
the equation of motion of Witten's string field theory:
\begin{equation}
Q_B \Psi_\lambda + \Psi_\lambda \ast \Psi_\lambda = 0 \,.
\end{equation}
The first piece of Schnabl's solution (\ref{Psi})
is $\Psi_\lambda$ with $\lambda=1$.
Recall that the second piece of (\ref{Psi})
was not involved in proving
that (\ref{Psi}) satisfies the equation of motion
contracted with any state in the Fock space
in \cite{Schnabl:2005gv}.
Solutions with other values of $\lambda$
were referred to as pure-gauge solutions in \cite{Schnabl:2005gv}.
It is not clear for what range of $\lambda$
the string field $\Psi_\lambda$ is well defined.
We derive (\ref{psi'_0-relation}) and (\ref{psi'_n-relation})
in various ways in this section.

\subsection{Solution in the CFT formulation}
\label{CFT-subsection}

Let us first prove
(\ref{psi'_0-relation}) and (\ref{psi'_n-relation})
in the CFT formulation.
The BRST transformations of
the operators $c (z)$ and $B$ are
\begin{equation}
Q_B \cdot c (z)
\equiv \oint \frac{d w}{2 \pi i} \, j_B (w) \, c(z)
= c \partial c (z) \,, \qquad
Q_B \cdot B = K \,,
\label{BRST-transformations}
\end{equation}
where $j_B (w)$ is the BRST current,
and the contour of the integral for $Q_B \cdot c (z)$
encircles $z$ counterclockwise.
The BRST transformations of the operators
$c \partial c (z)$ and $K$ vanish because $Q_B^2=0$.
It is then easy to show
(\ref{psi'_0-relation}) contracted with any state $\phi$
in the Fock space,
\begin{equation}
\langle \, \phi , Q_B \psi'_0 \, \rangle = 0 \,,
\end{equation}
using the expression
(\ref{psi'_0-inner-product-1}) or (\ref{psi'_0-inner-product-2}).

The inner product $\langle \, \phi , Q_B  \psi'_n \, \rangle$
with $n > 0$ can be derived from (\ref{psi'_n-CFT}) as follows:
\begin{eqnarray}
\langle \, \phi , Q_B  \psi'_n \, \rangle
&=& \left\langle \,
f_\infty \circ \phi (0) \,
c \partial c \left( \frac{\pi}{2} \right) \,
B \, K \, c \left( \frac{\pi \, (n+1)}{2} \right) \,
\right\rangle_{C_{\frac{\pi (n+2)}{2}}}
\nonumber \\
&& {}- \left\langle \,
f_\infty \circ \phi (0) \,
c \left( \frac{\pi}{2} \right) \,
K^2 \, c \left( \frac{\pi \, (n+1)}{2} \right) \,
\right\rangle_{C_{\frac{\pi (n+2)}{2}}}
\nonumber \\
&& {}+ \left\langle \,
f_\infty \circ \phi (0) \,
c \left( \frac{\pi}{2} \right) \,
B \, K \, c \partial c \left( \frac{\pi \, (n+1)}{2} \right) \,
\right\rangle_{C_{\frac{\pi (n+2)}{2}}} \,.
\end{eqnarray}
The inner product for the star product
$\langle \, \phi , \psi'_m \ast \psi'_{n-m} \, \rangle$
with $m, n-m > 0$ is given by
\begin{eqnarray}
&& \langle \, \phi , \psi'_m \ast \psi'_{n-m} \, \rangle
\nonumber \\
&=& \biggl\langle \,
f_\infty \circ \phi (0) \,
c \left( \frac{\pi}{2} \right) \, K \,
B \, c \left( \frac{\pi (m+1)}{2} \right) \,
c \left( \frac{\pi (m+2)}{2} \right) \, B \,
K \, c \left( \frac{\pi (n+2)}{2} \right) \,
\biggr\rangle_{C_{\frac{\pi (n+3)}{2}}} \,. \qquad
\label{star-product-example}
\end{eqnarray}
The first operator $f_\infty \circ \phi (0)$ is from
the string field $\phi$,
the next four operators are from $\psi'_m$,
and the last four operators are from $\psi'_{n-m}$.
Using (\ref{Bc+cB=1}) and $B^2=0$, we find
\begin{equation}
B \, c(z_1) \, c(z_2) \, B
= B \, c(z_1) - B \, c(z_1) \, B \, c(z_2)
= B \, c(z_1) - B \, c(z_2)
\label{B-c-c-B}
\end{equation}
for $z_1 < z_2$.
Then the inner product (\ref{star-product-example})
can be written as
\begin{eqnarray}
\langle \, \phi , \psi'_m \ast \psi'_{n-m} \, \rangle
&=& \biggl\langle \,
f_\infty \circ \phi (0) \,
c \left( \frac{\pi}{2} \right) \, K \, 
B \, c \left( \frac{\pi (m+1)}{2} \right) \,
K \, c \left( \frac{\pi (n+2)}{2} \right) \,
\biggr\rangle_{C_{\frac{\pi (n+3)}{2}}}
\nonumber \\
&& {}- \biggl\langle \,
f_\infty \circ \phi (0) \,
c \left( \frac{\pi}{2} \right) \, K \, 
B \, c \left( \frac{\pi (m+2)}{2} \right) \,
K \, c \left( \frac{\pi (n+2)}{2} \right) \,
\biggr\rangle_{C_{\frac{\pi (n+3)}{2}}} \,. \qquad
\end{eqnarray}
Note the simple dependence on $m$.
We therefore find
\begin{eqnarray}
&& \sum_{m=0}^{n} \,
\langle \, \phi , \psi'_m \ast \psi'_{n-m} \, \rangle
\nonumber \\
&=& \lim_{m \to 0} \biggl\langle \,
f_\infty \circ \phi (0) \,
c \left( \frac{\pi}{2} \right) \, K \, 
B \, c \left( \frac{\pi (m+1)}{2} \right) \,
K \, c \left( \frac{\pi (n+2)}{2} \right) \,
\biggr\rangle_{C_{\frac{\pi (n+3)}{2}}}
\nonumber \\
&& {}- \lim_{m \to n} \biggl\langle \,
f_\infty \circ \phi (0) \,
c \left( \frac{\pi}{2} \right) \, K \, 
B \, c \left( \frac{\pi (m+2)}{2} \right) \,
K \, c \left( \frac{\pi (n+2)}{2} \right) \,
\biggr\rangle_{C_{\frac{\pi (n+3)}{2}}} \,.
\end{eqnarray}
Using the formulas (\ref{c-K-c}) and (\ref{c-B-K-c-2}),
the sum can be written as
\begin{eqnarray}
\sum_{m=0}^{n} \,
\langle \, \phi , \psi'_m \ast \psi'_{n-m} \, \rangle
&=& {}- \biggl\langle \,
f_\infty \circ \phi (0) \,
c \partial c \left( \frac{\pi}{2} \right) \,
B \, K \, c \left( \frac{\pi (n+2)}{2} \right) \,
\biggr\rangle_{C_{\frac{\pi (n+3)}{2}}}
\nonumber \\
&& {}+ \biggl\langle \,
f_\infty \circ \phi (0) \,
c \left( \frac{\pi}{2} \right) \, K^2 \, 
c \left( \frac{\pi (n+2)}{2} \right) \,
\biggr\rangle_{C_{\frac{\pi (n+3)}{2}}}
\nonumber \\
&& {}- \biggl\langle \,
f_\infty \circ \phi (0) \,
c \left( \frac{\pi}{2} \right) \, B \, K \, 
c \partial c \left( \frac{\pi (n+2)}{2} \right) \,
\biggr\rangle_{C_{\frac{\pi (n+3)}{2}}} \,.
\end{eqnarray}
We have thus shown (\ref{psi'_n-relation})
contracted with any state $\phi$ in the Fock space:
\begin{equation}
\langle \, \phi , Q_B \psi'_{n+1} \, \rangle
= - \sum_{m=0}^{n} \,
\langle \, \phi , \psi'_m \ast \psi'_{n-m} \, \rangle
\end{equation}
for any integer $n$ with $n \ge 0$.

Before concluding the subsection, let us mention
a connection between the solution in the form (\ref{Psi})
and that in terms of the Bernoulli numbers (\ref{Psi-Bernoulli}).
Using the formula
\begin{equation}
\ket{n} \ast \ket{\phi}
= e^{\frac{\pi (n-1)}{2} K_1^L} \ket{\phi}
\label{n-ast-phi}
\end{equation}
for any string field $\phi$ \cite{Schnabl:2005gv, Schnabl:2002gg},
$\psi'_n$ with $n > 1$ can be written as
\begin{eqnarray}
\psi'_n &=& c_1 \ket{0} \ast \ket{n} \ast K_1^L B_1^L c_1 \ket{0}
= c_1 \ket{0} \ast \,
e^{\frac{\pi (n-1)}{2} K_1^L}
K_1^L B_1^L c_1 \ket{0}
\nonumber \\
&=& c_1 \ket{0} \ast \,
e^{-\frac{\pi}{4} K_1^L} e^{\frac{\pi n}{2} K_1^L}
K_1^L B_1^L e^{-\frac{\pi}{4} K_1^L} c_1 \ket{0}
\nonumber \\
&=& e^{\frac{\pi}{4} K_1^R} c_1 \ket{0} \ast \,
e^{\frac{\pi n}{2} K_1^L}
K_1^L B_1^L e^{-\frac{\pi}{4} K_1^L} c_1 \ket{0} \,.
\end{eqnarray}
The expression in the third line
is actually valid not only in the range $n > 1$
but also in the whole range $n > 0$.
The actions of the operators $e^{\frac{\pi}{4} K_1^R}$
and $e^{-\frac{\pi}{4} K_1^L}$ could be singular,
but it is not difficult to see from (\ref{psi'_n-CFT})
that the expression in the third line
is regular for the entire range $n > 0$,
and the limit $n \to 0$ is well defined.
The solution $\Psi_\lambda$ in (\ref{lambda})
can be formally written as
\begin{equation}
\Psi_\lambda = \sum_{n=0}^\infty \lambda^{n+1} \, \psi'_n
= \frac{2 \, \lambda}{\pi} \,
e^{\frac{\pi}{4} K_1^R} c_1 \ket{0} \ast \,
\frac{\frac{\pi}{2} K_1^L}
{1- \lambda \, e^{\frac{\pi}{2} K_1^L}} \,
B_1^L e^{-\frac{\pi}{4} K_1^L} c_1 \ket{0} \,.
\end{equation}
When $\lambda=1$, $\Psi_\lambda$ takes the form
\begin{equation}
\Psi_{\lambda=1} = {}- \frac{2}{\pi} \,
e^{\frac{\pi}{4} K_1^R} c_1 \ket{0} \ast \,
f_B \left( \frac{\pi}{2} K_1^L \right)
B_1^L e^{-\frac{\pi}{4} K_1^L} c_1 \ket{0} \,,
\end{equation}
where
\begin{equation}
f_B (x) = \frac{x}{e^x-1}
= \sum_{n=0}^\infty \frac{B_n \, x^n}{n!}
\end{equation}
is the generating function of the Bernoulli numbers $B_n$.
If we expand $1/(e^x-1)$ in $f_B (x)$ in powers of $e^x$,
the first piece of Schnabl's solution (\ref{Psi})
is reproduced.
If we instead expand the function $f_B (x)$ in powers of $x$,
the solution in terms of the Bernoulli numbers
(\ref{Psi-Bernoulli}) is obtained.
Recall the correspondence between
taking a derivative of $\psi_n$ with respect to $n$
and inserting an operator $K_1^L$ multiplied by $\pi/2$
we mentioned below (\ref{K_1^L-insertion}).

\subsection{Solution by an algebraic construction}
\label{algebraic-subsection}

It is also possible to prove
(\ref{psi'_0-relation}) and (\ref{psi'_n-relation})
in a more algebraic way.
We present such a proof in this subsection.

In this subsection, we define
\begin{eqnarray}
\psi'_0 &=& K_1^L c_1 \ket{0} + B_1^L c_0 c_1 \ket{0} \,,
\\
\psi'_n &=& c_1 \ket{0} \ast \ket{n} \ast B_1^L K_1^L c_1 \ket{0}
\end{eqnarray}
for integer $n$ in the range $n \ge 1$, where
\begin{equation}
\ket{n} = \underbrace{\ket{0} \ast \ket{0} \ast \ldots
\ast \ket{0}}_{n-1} \,,
\end{equation}
with the understanding that
\begin{equation}
\psi'_1 = c_1 \ket{0} \ast B_1^L K_1^L c_1 \ket{0} \,.
\end{equation}
In general, the state $\ket{n}$ with $n=1$ should be understood as
\begin{equation}
\ket{1} \ast \ket{\phi} = \ket{\phi}
\end{equation}
for any string field $\phi$.

Let us first list the equations we use to prove
(\ref{psi'_0-relation}) and (\ref{psi'_n-relation}):
\begin{eqnarray}
&& Q_B \, ( \phi_1 \ast \phi_2 ) = ( Q_B \, \phi_1 ) \ast \phi_2
+ (-1)^{\phi_1} \phi_1 \ast ( Q_B \, \phi_2 ) \,,
\phantom{\Big|}
\label{Q_B-derivation}
\\
&& Q_B^2 = 0 \,,
\phantom{\Big|}
\\
&& Q_B \ket{0} = 0 \,,
\phantom{\Big|}
\label{Q_B-vacuum}
\\
&& Q_B \, c_1 \ket{0} = {}- c_0 c_1 \ket{0} \,,
\phantom{\Big|}
\\
&& ( B_1^R \phi_1 ) \ast \phi_2
= {}- (-1)^{\phi_1} \phi_1 \ast ( B_1^L \phi_2 ) \,,
\phantom{\Big|}
\label{B-associativity}
\\
&& (B_1^L)^2 = (B_1^R)^2 = 0 \,,
\phantom{\Big|}
\label{B^2=0}
\\
&& ( B_1^L + B_1^R ) \ket{0} = 0 \,,
\phantom{\Big|}
\label{B-vacuum}
\\
&& ( B_1^L + B_1^R ) \, c_1 \ket{0} = \ket{0} \,,
\phantom{\Big|}
\label{B-c}
\\
&& \{ Q_B , B_1^L \} = K_1^L \,,
\phantom{\Big|}
\\
&& \{ Q_B , B_1^R \} = K_1^R \,,
\phantom{\Big|}
\label{K_1^R-equation}
\end{eqnarray}
for any string fields $\phi_1$ and $\phi_2$.
We also use the associativity of the star product.
It then follows from these equations that
for any string fields $\phi_1$ and $\phi_2$,
\begin{eqnarray}
&& ( K_1^R \, \phi_1 ) \ast \phi_2
= - \phi_1 \ast ( K_1^L \, \phi_2 ) \,,
\phantom{\Big|}
\\
&& [ Q_B , K_1^L ] = 0 \,,
\phantom{\Big|}
\\
&& [ B_1^L , K_1^L ] = 0 \,,
\phantom{\Big|}
\label{[B_1^L,K_1^L]}
\\
&& ( K_1^L + K_1^R ) \ket{0} = 0 \,,
\phantom{\Big|}
\\
&& B_1^L \ket{0} \ast \ket{0} = \ket{0} \ast B_1^L \ket{0} \,,
\phantom{\Big|}
\\
&& K_1^L \ket{0} \ast \ket{0} = \ket{0} \ast K_1^L \ket{0} \,.
\phantom{\Big|}
\label{[K_1^L,wedge]}
\end{eqnarray}
The string field $\psi'_n$ with $n \ge 1$ can be written as
\begin{equation}
\psi'_n = B_1^R c_1 \ket{0} \ast \ket{n} \ast K_1^L c_1 \ket{0}
\end{equation}
or as
\begin{equation}
\psi'_n =
- K_1^R c_1 \ket{0} \ast \ket{n} \ast B_1^L c_1 \ket{0} \,.
\end{equation}
The string field $\psi'_0$ is BRST exact:
\begin{equation}
\psi'_0 = K_1^L c_1 \ket{0} + B_1^L c_0 c_1 \ket{0}
= \{ Q_B , B_1^L \} \, c_1 \ket{0} - B_1^L Q_B \ket{0}
= Q_B \, B_1^L c_1 \ket{0} \,.
\label{BRST-exact}
\end{equation}
This proves (\ref{psi'_0-relation}).
We also use the following expression of $\psi'_0$:
\begin{equation}
\psi'_0 = Q_B \, B_1^L c_1 \ket{0}
= Q_B \ket{0} - Q_B \, B_1^R c_1 \ket{0}
= {}- Q_B \, B_1^R c_1 \ket{0}
= {}- K_1^R c_1 \ket{0} - B_1^R c_0 c_1 \ket{0} \,.
\end{equation}
The string field $Q_B \, \psi'_n$ with $n \ge 1$ is given by
\begin{equation}
Q_B \, \psi'_n
= {}- c_0 c_1 \ket{0} \ast \ket{n} \ast B_1^L K_1^L c_1 \ket{0}
- c_1 \ket{0} \ast \ket{n} \ast (K_1^L)^2 c_1 \ket{0}
- c_1 \ket{0} \ast \ket{n} \ast B_1^L K_1^L c_0 c_1 \ket{0} \,.
\label{Q_B-psi'_n-algebraic}
\end{equation}
The star product $\psi'_m \ast \psi'_{n-m}$
with $n, n-m \ge 1$ is given by
\begin{equation}
\psi'_m \ast \psi'_{n-m}
= - K_1^R c_1 \ket{0} \ast \ket{m} \ast B_1^L c_1 \ket{0}
\ast B_1^R c_1 \ket{0} \ast \ket{n-m} \ast K_1^L c_1 \ket{0} \,.
\end{equation}
Since
\begin{eqnarray}
B_1^L c_1 \ket{0} \ast B_1^R c_1 \ket{0}
&=& B_1^L c_1 \ket{0} \ast \ket{0}
- B_1^L c_1 \ket{0} \ast B_1^L c_1 \ket{0}
\phantom{\Big|} \nonumber \\
&=& B_1^L c_1 \ket{0} \ast \ket{0}
+ B_1^R B_1^L c_1 \ket{0} \ast c_1 \ket{0}
\phantom{\Big|} \nonumber \\
&=& B_1^L c_1 \ket{0} \ast \ket{0}
+ B_1^R \ket{0} \ast c_1 \ket{0}
- (B_1^R)^2 c_1 \ket{0} \ast c_1 \ket{0}
\phantom{\Big|} \nonumber \\
&=& B_1^L c_1 \ket{0} \ast \ket{0}
- \ket{0} \ast B_1^L c_1 \ket{0} \,,
\end{eqnarray}
the star product $\psi'_m \ast \psi'_{n-m}$
with $m, n-m \ge 1$ is given by
\begin{eqnarray}
\psi'_m \ast \psi'_{n-m}
&=& {}- K_1^R c_1 \ket{0} \ast \ket{m} \ast B_1^L c_1 \ket{0}
\ast \ket{n-m+1} \ast K_1^L c_1 \ket{0}
\nonumber \\
&& {}+ K_1^R c_1 \ket{0} \ast \ket{m+1} \ast B_1^L c_1 \ket{0}
\ast \ket{n-m} \ast K_1^L c_1 \ket{0} \,.
\end{eqnarray}
Note the simple dependence on $m$.
Therefore,
\begin{eqnarray}
\sum_{m=1}^{n-1} \psi'_m \ast \psi'_{n-m}
&=& {}- K_1^R c_1 \ket{0} \ast B_1^L c_1 \ket{0}
\ast \ket{n} \ast K_1^L c_1 \ket{0}
\nonumber \\
&& {}+ K_1^R c_1 \ket{0} \ast \ket{n} \ast B_1^L c_1 \ket{0}
\ast K_1^L c_1 \ket{0}
\end{eqnarray}
for $n \ge 2$.
The string product $\psi'_0 \ast \psi'_n$ with $n \ge 1$
is given by
\begin{eqnarray}
\psi'_0 \ast \psi'_n
&=& {}- K_1^R c_1 \ket{0}
\ast B_1^R c_1 \ket{0} \ast \ket{n} \ast K_1^L c_1 \ket{0}
- B_1^R c_0 c_1 \ket{0}
\ast B_1^R c_1 \ket{0} \ast \ket{n} \ast K_1^L c_1 \ket{0}
\phantom{\Big|} \nonumber \\
&=& {}- K_1^R c_1 \ket{0}
\ast \ket{n+1} \ast K_1^L c_1 \ket{0}
+ K_1^R c_1 \ket{0}
\ast B_1^L c_1 \ket{0} \ast \ket{n} \ast K_1^L c_1 \ket{0}
\phantom{\Big|} \nonumber \\
&& {}- B_1^R c_0 c_1 \ket{0}
\ast \ket{n+1} \ast K_1^L c_1 \ket{0}
+ B_1^R c_0 c_1 \ket{0}
\ast B_1^L c_1 \ket{0} \ast \ket{n} \ast K_1^L c_1 \ket{0}
\phantom{\Big|} \nonumber \\
&=& c_1 \ket{0}
\ast \ket{n+1} \ast (K_1^L)^2 c_1 \ket{0}
+ K_1^R c_1 \ket{0}
\ast B_1^L c_1 \ket{0} \ast \ket{n} \ast K_1^L c_1 \ket{0}
\phantom{\Big|} \nonumber \\
&& {}+ c_0 c_1 \ket{0}
\ast \ket{n+1} \ast B_1^L K_1^L c_1 \ket{0}
- c_0 c_1 \ket{0}
\ast (B_1^L)^2 c_1 \ket{0} \ast \ket{n} \ast K_1^L c_1 \ket{0}
\phantom{\Big|} \nonumber \\
&=& K_1^R c_1 \ket{0}
\ast B_1^L c_1 \ket{0} \ast \ket{n} \ast K_1^L c_1 \ket{0}
\phantom{\Big|} \nonumber \\
&& {}+ c_0 c_1 \ket{0}
\ast \ket{n+1} \ast B_1^L K_1^L c_1 \ket{0}
+ c_1 \ket{0}
\ast \ket{n+1} \ast (K_1^L)^2 c_1 \ket{0} \,.
\end{eqnarray}
The string field $\psi'_n \ast \psi'_0$ with $n \ge 1$ is given by
\begin{eqnarray}
\psi'_n \ast \psi'_0
&=& {}- K_1^R c_1 \ket{0} \ast \ket{n} \ast B_1^L c_1 \ket{0}
\ast K_1^L c_1 \ket{0}
- K_1^R c_1 \ket{0} \ast \ket{n} \ast B_1^L c_1 \ket{0}
\ast B_1^L c_0 c_1 \ket{0}
\phantom{\Big|} \nonumber \\
&=& {}- K_1^R c_1 \ket{0} \ast \ket{n} \ast B_1^L c_1 \ket{0}
\ast K_1^L c_1 \ket{0}
\phantom{\Big|} \nonumber \\
&& {}- K_1^R c_1 \ket{0} \ast \ket{n+1} \ast B_1^L c_0 c_1 \ket{0}
+ K_1^R c_1 \ket{0} \ast \ket{n} \ast B_1^R c_1 \ket{0}
\ast B_1^L c_0 c_1 \ket{0}
\phantom{\Big|} \nonumber \\
&=& {}- K_1^R c_1 \ket{0} \ast \ket{n} \ast B_1^L c_1 \ket{0}
\ast K_1^L c_1 \ket{0}
\phantom{\Big|} \nonumber \\
&& {}- K_1^R c_1 \ket{0} \ast \ket{n+1} \ast B_1^L c_0 c_1 \ket{0}
+ K_1^R c_1 \ket{0} \ast \ket{n} \ast c_1 \ket{0}
\ast (B_1^L)^2 c_0 c_1 \ket{0}
\phantom{\Big|} \nonumber \\
&=& {}- K_1^R c_1 \ket{0} \ast \ket{n} \ast B_1^L c_1 \ket{0}
\ast K_1^L c_1 \ket{0}
+ c_1 \ket{0} \ast \ket{n+1} \ast B_1^L K_1^L c_0 c_1 \ket{0} \,.
\end{eqnarray}
Therefore, we find
\begin{eqnarray}
\sum_{m=0}^{n} \psi'_m \ast \psi'_{n-m}
&=& c_0 c_1 \ket{0}
\ast \ket{n+1} \ast B_1^L K_1^L c_1 \ket{0}
+ c_1 \ket{0}
\ast \ket{n+1} \ast (K_1^L)^2 c_1 \ket{0}
\nonumber \\
&& {}+ c_1 \ket{0} \ast \ket{n+1} \ast B_1^L K_1^L c_0 c_1 \ket{0}
\end{eqnarray}
for $n \ge 1$.
We have thus shown (\ref{psi'_n-relation}) for $n \ge 1$
by combining this with (\ref{Q_B-psi'_n-algebraic}).
Finally, the string product $\psi'_0 \ast \psi'_0$ is given by
\begin{eqnarray}
\psi'_0 \ast \psi'_0
&=& {}- K_1^R c_1 \ket{0} \ast K_1^L c_1 \ket{0}
- K_1^R c_1 \ket{0} \ast B_1^L c_0 c_1 \ket{0}
\phantom{\Big|} \nonumber \\
&& {}- B_1^R c_0 c_1 \ket{0} \ast K_1^L c_1 \ket{0}
- B_1^R c_0 c_1 \ket{0} \ast B_1^L c_0 c_1 \ket{0}
\phantom{\Big|} \nonumber \\
&=& c_1 \ket{0} \ast (K_1^L)^2 c_1 \ket{0}
+ c_1 \ket{0} \ast B_1^L K_1^L c_0 c_1 \ket{0}
\phantom{\Big|} \nonumber \\
&& {}+ c_0 c_1 \ket{0} \ast B_1^L K_1^L c_1 \ket{0}
+ c_0 c_1 \ket{0} \ast (B_1^L)^2 c_0 c_1 \ket{0}
\phantom{\Big|} \nonumber \\
&=& c_0 c_1 \ket{0} \ast B_1^L K_1^L c_1 \ket{0}
+ c_1 \ket{0} \ast (K_1^L)^2 c_1 \ket{0}
+ c_1 \ket{0} \ast B_1^L K_1^L c_0 c_1 \ket{0} \,,
\end{eqnarray}
and thus $Q_B \, \psi'_1 = - \psi'_0 \ast \psi'_0$.
This completes the proof of (\ref{psi'_n-relation})
for the whole range $n \ge 0$.

We have demonstrated that
(\ref{psi'_0-relation}) and (\ref{psi'_n-relation})
can be shown from the set of
the equations from (\ref{Q_B-derivation})
to (\ref{K_1^R-equation}).
The first two equations are
general properties of the BRST operator
and requisites for string field theory
to make sense.
The operator $c_0$ only appears in the fourth equation
and can be eliminated if we replace ${}- c_0 c_1 \ket{0}$
by $Q_B \, c_1 \ket{0}$.
Similarly, $K_1^L$ and $K_1^R$ can also be eliminated
if we replace them by $\{ Q_B , B_1^L \}$ and $\{ Q_B , B_1^R \}$,
respectively.
Now the solution and the proof can be written
in terms of $\ket{0}$, $Q_B$, $B_1^L$, $B_1^R$, and $c_1$,
and the equations we need are
(\ref{Q_B-vacuum}), (\ref{B-associativity}),
(\ref{B^2=0}), (\ref{B-vacuum}), and (\ref{B-c}).
We can construct a different solution
if we find a different set of these ingredients
which also satisfy the reduced set of the equations
shown above.
A rather trivial class of replacements is
$\ket{0} \to e^A \ket{0}$,
$B_1^L \to e^A B_1^L e^{-A}$,
$B_1^R \to e^A B_1^R e^{-A}$,
and $c_1 \to e^A c_1 e^{-A}$,
where $A$ is a linear combination of
$K_n = L_n - (-1)^n L_{-n}$.
Since $A$ is a derivation of the star product
and commutes with $Q_B$,
the solution $\Psi_\lambda$ is replaced by
$e^A \, \Psi_\lambda$, which obviously satisfies
the equation of motion.
It also seems possible to replace only $B_1^L$ and $B_1^R$
by different half integrals of the $b$ ghost.
There will be many other ways to modify the solutions,
while the reduced set of equations are satisfied.
However, it is not clear if we can obtain inequivalent solutions
in this way.

\subsection{Solution in the half-string picture}
\label{half-string-subsection}

Schnabl's solution can be naturally described
in the half-string picture,
where a string field is considered
as an operator acting on the Hilbert space of the half string
\cite{Witten:1985cc, Chan:1986ie, Chan:1988px, Bordes:1989xh,
Abdurrahman:1993ip, Abdurrahman:1994ix, Abdurrahman:1998gu,
Rastelli:2001rj, Gross:2001rk, Kawano:2001fn, Gross:2001yk,
Furuuchi:2001df, Moeller:2001ap, Erler:2003eq, Fuchs:2003wu}.
Witten's star product then corresponds
to the multiplication of the operators.
There is a subtle issue
on how to deal with the open string midpoint in this formalism,
but we do not attempt to make it rigorous here.
We would rather use the formalism
to help gain more intuition
in manipulating the solution.
It is always possible to translate the insight
we get in this half-string picture
into other languages in previous subsections.

Let us denote the operator
associated with the vacuum state $\ket{0}$
by $e^{\frac{\pi}{2} K}$:
\begin{equation}
\ket{0} \, \sim \, e^{\frac{\pi}{2} K} \,.
\end{equation}
This can be thought of as a definition of $K$.
The wedge state $\ket{n}$ then corresponds to
\begin{equation}
\ket{n} \sim \, e^{\frac{\pi (n-1)}{2} K} \,.
\end{equation}
The notation is motivated by the formula (\ref{n-ast-phi}).
In fact, the operator $K$ here corresponds
to an insertion of $K$ defined in (\ref{K-definition}).
The state $c_1 \ket{0}$ corresponds to $c(0)$
in the state-operator mapping.
The upper half of the unit disk with the insertion
$c(0)$ at the origin in the coordinate $\xi$ is mapped
by the conformal transformation
$z = f_\infty (\xi) = \arctan \xi$
to a semi-infinite strip with a width of $\pi/2$
which has an insertion of $c(0)$ at the origin.
Each of the right and left halves of the strip
corresponds to $e^{\frac{\pi}{4} K}$.
The state $c_1 \ket{0}$ can therefore be expressed as
\begin{equation}
c_1 \ket{0} \,
\sim \, e^{\frac{\pi}{4} K} \, c \, e^{\frac{\pi}{4} K} \,,
\end{equation}
where $c$ corresponds to a $c$-ghost insertion
at the end of the half string.

If we denote the operator
corresponding to the string field $\ket{\phi}$
by $\Phi$, the actions of $B_1^L$ and $B_1^R$
can be written using a Grassmann-odd operator $B$ as
\begin{equation}
B_1^L \ket{\phi} \, \sim \, B \, \Phi \,, \qquad
B_1^R \ket{\phi} \, \sim \, {}-(-1)^\Phi \Phi \, B \,,
\end{equation}
where $(-1)^{\Phi}=1$ when $\Phi$ is Grassmann even
and $(-1)^{\Phi}=-1$ when $\Phi$ is Grassmann odd.
The operator $B$ here corresponds
to an insertion of $B$ defined in (\ref{B-definition}).
The relation (\ref{B-associativity})
is nothing but the associativity of the operator multiplication
in the half-string picture.
Similarly,
\begin{equation}
K_1^L \ket{\phi} \, \sim \, K \, \Phi \,, \qquad
K_1^R \ket{\phi} \, \sim \, {}-\Phi \, K \,.
\end{equation}
Note that the operator $K$ here coincides
with the one appeared in the operator corresponding to $\ket{0}$.

We have introduced the operators $K$, $B$, and $c$
without providing their detailed definitions.
All we need is the following set of commutation relations:
\begin{equation}
{}[ \,  K , B \, ] = 0 \,, \qquad
{}\{ \, B , c \, \} = 1 \,, \qquad
c^2 = 0 \,, \qquad
B^2 = 0 \,.
\end{equation}
Note that $[ \, K , c \, ] \ne 0$.

The action of the BRST operator $Q_B$ is a derivation
with respect to the operator multiplication.
Let us denote it by $d$:
\begin{equation}
Q_B \ket{\phi} \, \sim \, d \, \Phi \,.
\end{equation}
The derivation property of the Grassmann-odd operation $d$
can be expressed as
\begin{equation}
d \, ( \, \Phi_1 \, \Phi_2 \, )
= ( \, d \, \Phi_1 \, ) \, \Phi_2
+ (-1)^{\Phi_1} \Phi_1 \, ( \, d \, \Phi_2 \, )
\end{equation}
for any pair of operators $\Phi_1$ and $\Phi_2$,
where $(-1)^{\Phi_1}=1$ when $\Phi_1$ is Grassmann even
and $(-1)^{\Phi_1}=-1$ when $\Phi_1$ is Grassmann odd.
The actions of $d$ on $B$, $K$, and $c$ are given by
\begin{equation}
d \, B = K \,, \qquad
d \, K = 0 \,, \qquad
d \, c = c \, K \, c \,.
\end{equation}
The last equation follows from (\ref{c-K-c})
and (\ref{BRST-transformations}).
It is easy to verify that $d^2=0$
for any operator made of $K$, $B$, and $c$.

The string field $\psi'_n$ corresponds to
\begin{equation}
\psi'_n \, \sim \,
e^{\frac{\pi}{4} K} \, c \,
B \, K \, e^{\frac{\pi \, n}{2} K} \,
c \, e^{\frac{\pi}{4} K} \,.
\end{equation}
This can be easily seen from figure \ref{figure3}.
The solution $\Psi_\lambda$ in (\ref{lambda})
can be formally written as
\begin{equation}
\Psi_\lambda \, \sim \,
\lambda \, e^{\frac{\pi}{4} K} \, c \,
\frac{B \, K}{1- \lambda \, e^{\frac{\pi}{2} K}} \,
c \, e^{\frac{\pi}{4} K}
= f(K) \, c \,
\frac{B \, K}{1- f(K)^2} \,
c \, f(K) \,,
\end{equation}
where
\begin{equation}
f(K) = \sqrt{\lambda} \, e^{\frac{\pi}{4} K} \,.
\end{equation}
Let us prove that
$Q_B \Psi_\lambda + \Psi_\lambda \ast \Psi_\lambda = 0$
in the half-string picture.
The string field  $Q_B \Psi_\lambda$ corresponds to
\begin{eqnarray}
Q_B \Psi_\lambda &\sim&
d \left[ \, f(K) \, c \,
\frac{B \, K}{1- f(K)^2} \,
c \, f(K) \, \right]
\nonumber \\
&=& f(K) \, c \, K \, c \,
\frac{B \, K}{1- f(K)^2} \,
c \, f(K)
- f(K) \, c \,
\frac{K^2}{1- f(K)^2} \,
c \, f(K)
+ f(K) \, c \,
\frac{B \, K}{1- f(K)^2} \,
c \, K \, c \, f(K) \,.
\nonumber \\
\end{eqnarray}
The string field $\Psi_\lambda \ast \Psi_\lambda$ corresponds to
\begin{eqnarray}
\Psi_\lambda \ast \Psi_\lambda &\sim&
f(K) \, c \, \frac{B \, K}{1- f(K)^2} \,
c \, f(K)^2 \, c \, \frac{B \, K}{1- f(K)^2} \, c \, f(K)
\nonumber \\
&=& f(K) \, c \, K \, \frac{1}{1- f(K)^2} \,
B \, c \, f(K)^2 \, c \, B \,
\frac{1}{1- f(K)^2} \, K \, c \, f(K) \,.
\end{eqnarray}
Since
\begin{eqnarray}
&& B \, c \, f(K)^2 \, c \, B
= B \, c \, f(K)^2 \, ( 1 - B \, c )
= B \, c \, f(K)^2 - B \, c \, B \, f(K)^2 \, c
\nonumber \\
&=& B \, c \, f(K)^2 - B \, f(K)^2 \, c
= B \, c \, f(K)^2 - f(K)^2 \, B \, c
= [ \, B \, c \,,\, f(K)^2 \, ] \,,
\end{eqnarray}
$\Psi_\lambda \ast \Psi_\lambda$ is given by
\begin{eqnarray}
\Psi_\lambda \ast \Psi_\lambda &\sim&
f(K) \, c \, K \, \frac{1}{1- f(K)^2} \,
[ \, B \, c \,,\, f(K)^2 \, ] \,
\frac{1}{1- f(K)^2} \, K \, c \, f(K)
\nonumber \\
&=& f(K) \, c \, K \,
[ \, B \, c \,,\, \frac{1}{1- f(K)^2} \, ] \,
K \, c \, f(K)
\nonumber \\
&=& f(K) \, c \, K \, B \, c \, \frac{1}{1- f(K)^2} \,
K \, c \, f(K)
- f(K) \, c \, K \, \frac{1}{1- f(K)^2} B \, c \,
K \, c \, f(K)
\nonumber \\
&=& {}- f(K) \, c \, K \, c \, \frac{B \, K}{1- f(K)^2} \, c \, f(K)
+ f(K) \, c \, \frac{K^2}{1- f(K)^2} \, c \, f(K)
\nonumber \\
&& {}- f(K) \, c \, \frac{B \, K}{1- f(K)^2} \,
c \, K \, c \, f(K) \,.
\end{eqnarray}
We have thus shown that
$Q_B \Psi_\lambda + \Psi_\lambda \ast \Psi_\lambda = 0$.
If we expand $1/(1- f(K)^2)$ in powers of $e^{\frac{\pi}{2} K}$,
the proof in subsection \ref{CFT-subsection} is reproduced.
If we instead expand $K/(1- f(K)^2)$ in powers of $K$
when $\lambda=1$,
it will formally give a proof
that the solution written in terms of the Bernoulli numbers
(\ref{Psi-Bernoulli}) satisfies the equation of motion.
However, the expansion in the form of (\ref{Psi-Bernoulli})
will not converge
so it is not clear whether the proof for this form is useful.

The proof can be further simplified and made more symmetric
if we note
\begin{eqnarray}
B \, c \, f(K)^2 \, c \, B
&=& {}- B \, c \, ( \, 1 - f(K)^2 \, ) \, c \, B
= {}- ( 1 - c \, B ) \, ( \, 1 - f(K)^2 \, ) \, ( 1 - B \, c )
\nonumber \\
&=& {}- ( \, 1 - f(K)^2 \, )
+ ( \, 1 - f(K)^2 \, ) \, B \, c
+ c \, B \, ( \, 1 - f(K)^2 \, )
\nonumber \\
&=& {}- ( \, 1 - f(K)^2 \, )
+ ( \, 1 - f(K)^2 \, ) \, ( 1 - c \, B )
+ ( 1 - B \, c ) \, ( \, 1 - f(K)^2 \, )
\nonumber \\
&=& {} ( \, 1 - f(K)^2 \, )
- ( \, 1 - f(K)^2 \, ) \, c \, B
- B \, c \, ( \, 1 - f(K)^2 \, ) \,.
\end{eqnarray}
We then immediately obtain
\begin{eqnarray}
\Psi_\lambda \ast \Psi_\lambda &\sim&
f(K) \, c \, K \, \frac{1}{1- f(K)^2} \,
B \, c \, f(K)^2 \, c \, B \,
\frac{1}{1- f(K)^2} \, K \, c \, f(K)
\nonumber \\
&=& f(K) \, c \, \frac{K^2}{1- f(K)^2} \, c \, f(K)
- f(K) \, c \, K \, c \, \frac{B \, K}{1- f(K)^2} \, c \, f(K)
\nonumber \\
&& {}- f(K) \, c \, \frac{B \, K}{1- f(K)^2} \,
c \, K \, c \, f(K) \,.
\end{eqnarray}
Note that we have never used the explicit form of $f(K)$.
Therefore, we can formally construct a solution
for any choice of the function $f(K)$.
It is an important open problem
to understand when solutions are well defined
and when they become inequivalent.

\subsection{Solution as a pure-gauge configuration}
\label{pure-gauge-subsection}

The first piece of Schnabl's solution $\Psi$ in (\ref{Psi})
can be formally written as a pure-gauge configuration
$e^{-\Lambda} \, ( Q_B \, e^{\Lambda} \, )$
with some gauge parameter $\Lambda$.
Here and in what follows in this subsection
products of string fields are defined using the star product
even when the star symbol is omitted.

As we have seen in (\ref{BRST-exact}),
the string field $\psi'_0$ is BRST exact:
\begin{equation}
\psi'_0 = Q_B \, \Phi \,,
\end{equation}
where
\begin{equation}
\Phi = B_1^L \, c_1 \ket{0} \,.
\end{equation}
A crucial observation is that
the string field $\psi'_n$ for integer $n$ with $n \ge 1$
can also be written in terms of $Q_B$ and $\Phi$:
\begin{equation}
\psi'_n = ( Q_B \, \Phi ) \, \Phi^n \,.
\label{psi'_n-Q_B-Phi}
\end{equation}
It is easy to see in the CFT formulation
that $\psi'_n = \psi'_0 \, \Phi^n$
by repeatedly using the relation $B \, c(z) \, B = B$.
It can also be shown in the following way.
Using (\ref{B-c}), (\ref{B-associativity}),
and (\ref{B^2=0}), we find that
\begin{equation}
\Phi^2 = \ket{0} \ast \Phi \,.
\end{equation}
Therefore,
\begin{equation}
\Phi^n = \underbrace{\ket{0} \ast \ket{0} \ast \ldots
\ast \ket{0}}_{n-1} \ast \, \Phi
= \ket{n} \ast \Phi \,.
\end{equation}
Since
\begin{equation}
Q_B \, \Phi = \psi'_0
= {}- K_1^R c_1 \ket{0} - B_1^R c_0 c_1 \ket{0} \,,
\end{equation}
we obtain
\begin{equation}
( Q_B \, \Phi ) \, \Phi^n
= {}- K_1^R c_1 \ket{0} \ast \ket{n} \ast B_1^L \, c_1 \ket{0}
= c_1 \ket{0} \ast \ket{n} \ast B_1^L \, K_1^L \, c_1 \ket{0}
= \psi'_n \,.
\end{equation}
The solution $\Psi_\lambda$ can be written as
\begin{equation}
\Psi_\lambda = \sum_{n=0}^\infty \lambda^{n+1} \, \psi'_n
= \sum_{n=0}^\infty \lambda^{n+1} \, ( Q_B \, \Phi ) \, \Phi^n
= \lambda \, ( Q_B \, \Phi ) \, \frac{1}{1- \lambda \, \Phi} \,.
\end{equation}
Since
$e^{-\Lambda} \, ( Q_B \, e^{\Lambda} \, )
= - ( Q_B \, e^{-\Lambda} \, ) \, e^{\Lambda}$,
the gauge parameter $\Lambda$ can be written in terms of $\Phi$
as follows:
\begin{equation}
e^{\Lambda} = \frac{1}{1- \lambda \, \Phi} \,,
\end{equation}
or
\begin{equation}
\Lambda = -\ln \, (1- \lambda \, \Phi)
= \sum_{n=1}^\infty \frac{\lambda^n}{n} \Phi^n \,.
\end{equation}
It is now straightforward to see that $\Psi_\lambda$ satisfies
the equation of motion.
Since
\begin{equation}
Q_B \frac{1}{1-\lambda \, \Phi}
= \frac{1}{1-\lambda \, \Phi} \, \lambda \, ( Q_B \, \Phi ) \,
\frac{1}{1-\lambda \, \Phi} \,,
\end{equation}
$Q_B \, \Psi_\lambda$ is given by
\begin{equation}
Q_B \, \Psi_\lambda
= {}- \lambda \, ( Q_B \, \Phi ) \, Q_B \frac{1}{1- \lambda \, \Phi}
= {}- \lambda \, ( Q_B \, \Phi ) \, \frac{1}{1-\lambda \, \Phi} \,
\lambda \, ( Q_B \, \Phi ) \,
\frac{1}{1-\lambda \, \Phi} = - \Psi_\lambda^2 \,.
\end{equation}
Therefore,
\begin{equation}
Q_B \, \Psi_\lambda + \Psi_\lambda^2 = 0 \,.
\end{equation}
It is also straightforward to calculate $Q_B \, \psi'_n$
using the expression (\ref{psi'_n-Q_B-Phi}):
\begin{eqnarray}
Q_B \, \psi'_n &=& Q_B \, [ \, ( Q_B \, \Phi ) \, \Phi^n \, ]
= {}- ( Q_B \, \Phi ) \, ( Q_B \, \Phi^n )
= {}- \sum_{m=0}^{n-1} ( Q_B \, \Phi ) \,
\Phi^m \, ( Q_B \, \Phi ) \, \Phi^{n-m-1}
\nonumber \\
&=& {}- \sum_{m=0}^{n-1} \psi'_m \, \psi'_{n-m-1}
\end{eqnarray}
for $n \ge 1$. We have thus reproduced (\ref{psi'_n-relation}).

The string field $\psi'_0$ can also be written as
\begin{equation}
\psi'_0 = Q_B \, \widetilde{\Phi} \,,
\end{equation}
where
\begin{equation}
\widetilde{\Phi} = {}- B_1^R c_1 \ket{0} \,.
\end{equation}
The string field $\psi'_n$ for integer $n$ with $n \ge 1$
can also be written in terms of $Q_B$ and $\widetilde{\Phi}$:
\begin{equation}
\psi'_n = (-1)^n \, \widetilde{\Phi}^n \,
( Q_B \, \widetilde{\Phi} ) \,.
\end{equation}
More generally, it can be written
using both $\Phi$ and $\widetilde{\Phi}$ as
\begin{equation}
\psi'_n = (-1)^m \, \widetilde{\Phi}^m \,
( Q_B \, \Phi ) \, \Phi^{n-m}
= (-1)^m \, \widetilde{\Phi}^m \,
( Q_B \, \widetilde{\Phi} ) \, \Phi^{n-m} \,,
\end{equation}
for any integer $m$ with $0 \le m \le n$.

\section{Kinetic term}
\label{Kinetic-term-section}
\setcounter{equation}{0}

The kinetic term of the Witten's string field theory action
was evaluated for Schnabl's solution in \cite{Schnabl:2005gv}:
\begin{eqnarray}
{\cal K}_2
&=& \lim_{N \to \infty} \sum_{n=0}^N \sum_{m=0}^N \,
\langle \, \psi'_n , Q_B \psi'_m \, \rangle
= \frac{1}{2} - \frac{1}{\pi^2} \,,
\\
{\cal K}_1
&=& \lim_{N \to \infty} \sum_{m=0}^N \,
\langle \, \psi_N , Q_B \psi'_m \, \rangle
= \frac{1}{2} + \frac{2}{\pi^2} \,,
\\
{\cal K}_0
&=& \lim_{N \to \infty}
\langle \, \psi_N , Q_B \psi_N \, \rangle
= \frac{1}{2} + \frac{2}{\pi^2} \,.
\end{eqnarray}
The subscript of ${\cal K}$ indicates the number of sums.
We reproduce these results in a different way
in this section.
Our way of calculating these quantities
makes it clear how the $\psi_N$ piece cancels
the difference between ${\cal K}_2$
and the value (\ref{prediction-kinetic})
predicted by Sen's conjecture.
Our method also has an advantage
in generalizing to the calculations for the cubic term
in the next section.

The inner product
$\langle \, \psi_n , Q_B \psi_m \, \rangle$
was calculated in \cite{Schnabl:2005gv}:
\begin{eqnarray}
&& \langle \, \psi_n , Q_B \psi_m \, \rangle
\nonumber \\
&=& \frac{1}{\pi^2} \,
\left( 1 + \cos \frac{\pi \, (m-n)}{m+n+2} \right)
\left( -1 + \frac{m+n+2}{\pi} \sin \frac{2 \, \pi}{m+n+2} \right)
\nonumber \\
&& {}+ 2 \sin^2 \frac{\pi}{m+n+2}
\left[ \, -\frac{m+n+1}{\pi^2}
+ \frac{m n}{\pi^2} \cos \frac{\pi \, (m-n)}{m+n+2}
+ \frac{(m+n+2)(m-n)}{2 \, \pi^3} \sin \frac{\pi \, (m-n)}{m+n+2} \,
\right] \,.
\nonumber \\
\end{eqnarray}
Using this expression, it was shown in \cite{Schnabl:2005gv} that
\begin{equation}
\sum_{m=0}^n \langle \, \psi'_m , Q_B \psi'_{n-m} \, \rangle = 0 \,.
\label{psi'-Q_B-psi'}
\end{equation}
The double sum in ${\cal K}_2$
before taking the limit $N \to \infty$
can be decomposed in the following way:
\begin{eqnarray}
\sum_{n=0}^N \sum_{m=0}^N \,
\langle \, \psi'_n , Q_B \psi'_m \, \rangle
&=& \sum_{m=0}^N \sum_{n=0}^{N-m} \,
\langle \, \psi'_n , Q_B \psi'_m \, \rangle
+ \sum_{m=1}^{N} \sum_{n=N-m+1}^N \,
\langle \, \psi'_n , Q_B \psi'_m \, \rangle
\nonumber \\
&=& \sum_{n=0}^N \sum_{m=0}^n \,
\langle \, \psi'_m , Q_B \psi'_{n-m} \, \rangle
+ \sum_{m=1}^{N} \sum_{n=N-m+1}^N \,
\langle \, \psi'_n , Q_B \psi'_m \, \rangle \,.
\end{eqnarray}
The first double sum in the last line vanishes
because of (\ref{psi'-Q_B-psi'}).
Therefore, ${\cal K}_2$ reduces to
\begin{equation}
{\cal K}_2
= \lim_{N \to \infty} \sum_{m=1}^{N} \sum_{n=N-m+1}^N \,
\langle \, \psi'_n , Q_B \psi'_m \, \rangle \,.
\label{K_2-reduced}
\end{equation}
Now all of ${\cal K}_2$, ${\cal K}_1$, and ${\cal K}_0$
are written in terms of inner products of the form
$\langle \, \psi_n , Q_B \psi_m \, \rangle$
with large $n+m$ and their derivatives.
When $n+m$ is large, the inner product
$\langle \, \psi_n , Q_B \psi_m \, \rangle$ becomes
\begin{eqnarray}
&& \lim_{n+m \to \infty}
\langle \, \psi_n , Q_B \psi_m \, \rangle
\nonumber \\
&=& \frac{1}{\pi^2} \,
\left( 1 + \cos \frac{\pi \, (m-n)}{m+n} \right)
+ \frac{2 \, m n}{(m+n)^2}
\cos \frac{\pi \, (m-n)}{m+n}
+ \frac{m-n}{\pi (m+n)} \sin \frac{\pi \, (m-n)}{m+n} \,.
\label{n+m-large}
\end{eqnarray}
Note that only the sum $n+m$ needs to be large
for this expression to be valid,
and either $n$ or $m$ can be small as long as the sum is large.
Let us introduce the function
\begin{equation}
{\cal F}_K (x,y) = \frac{1}{\pi^2} \,
\left( 1 + \cos \frac{\pi \, (x-y)}{x+y} \right)
+ \frac{2 \, x y}{(x+y)^2}
\cos \frac{\pi \, (x-y)}{x+y}
+ \frac{x-y}{\pi (x+y)} \sin \frac{\pi \, (x-y)}{x+y} \,.
\end{equation}
Note that
\begin{equation}
{\cal F}_K (0,y) = {\cal F}_K (x,0) = 0 \,,
\end{equation}
and
\begin{equation}
{\cal F}_K (ax, ay) = {\cal F}_K (x,y)
\label{F_K-identity}
\end{equation}
for any nonvanishing $a$.
The inner product (\ref{n+m-large}) in the limit can be written as
\begin{equation}
\lim_{n+m \to \infty}
\langle \, \psi_n , Q_B \psi_m \, \rangle
= {\cal F}_K \left( a n, a m \right)
\end{equation}
for any nonvanishing $a$. In particular, we can choose
$a$ to be $1/N$ in taking the limit $N \to \infty$:
\begin{equation}
\lim_{n+m \to \infty}
\langle \, \psi_n , Q_B \psi_m \, \rangle
= {\cal F}_K \left( \frac{n}{N}, \frac{m}{N} \right) \,.
\end{equation}
The quantity ${\cal K}_0$ is simply given by
\begin{equation}
{\cal K}_0
= \lim_{N \to \infty} \langle \, \psi_N , Q_B \psi_N \, \rangle
= {\cal F}_K (1,1) = \frac{2}{\pi^2} + \frac{1}{2} \,.
\end{equation}
The quantity ${\cal K}_1$ can be written as
\begin{equation}
{\cal K}_1
= \lim_{N \to \infty} \sum_{n=0}^N \,
\langle \, \psi'_n , Q_B \psi_N \, \rangle
= \lim_{N \to \infty} \sum_{n=0}^N \, \frac{\partial}{\partial n}
{\cal F}_K \left( \frac{n}{N}, 1 \right)
= \lim_{N \to \infty} \frac{1}{N} \sum_{n=0}^N \,
\frac{\partial}{\partial x}
{\cal F}_K \left( x, 1 \right) \biggr|_{x=\frac{n}{N}} \,.
\end{equation}
Using the formula
\begin{equation}
\lim_{N \to \infty} \frac{1}{N} \, \sum_{n=0}^N
f \left( \frac{n}{N} \right)
= \int_0^1 dx \, f(x) \,,
\end{equation}
we obtain
\begin{equation}
{\cal K}_1
= \int_0^1 dx \, \frac{\partial}{\partial x}
{\cal F}_K \left( x, 1 \right)
= {\cal F}_K (1,1) - {\cal F}_K (0,1)
= {\cal F}_K (1,1) = \frac{2}{\pi^2} + \frac{1}{2} \,,
\end{equation}
where we used that ${\cal F}_K (0,y) = 0$.
It is more or less obvious in this way of the calculation
that ${\cal K}_1$ and ${\cal K}_0$ coincide,
while it was not the case in \cite{Schnabl:2005gv},
where ${\cal K}_1$ was obtained from a nontrivial integral.

The expression of ${\cal K}_2$ in (\ref{K_2-reduced})
can also be transformed into an integral:
\begin{eqnarray}
{\cal K}_2
&=& \lim_{N \to \infty} \sum_{m=1}^{N} \sum_{n=N-m+1}^N \,
\frac{\partial}{\partial n} \frac{\partial}{\partial m}
{\cal F}_K \left( \frac{n}{N}, \frac{m}{N} \right)
\nonumber \\
&=& \lim_{N \to \infty} \frac{1}{N^2}
\sum_{m=1}^{N} \sum_{n=N-m+1}^N \,
\frac{\partial}{\partial x} \frac{\partial}{\partial y}
{\cal F}_K \left( x, y \right)
\biggr|_{x=\frac{n}{N} ,\, y=\frac{m}{N}}
= \int_0^1 dy \int_{1-y}^1 dx \,
\frac{\partial}{\partial x} \frac{\partial}{\partial y}
{\cal F}_K \left( x, y \right) \,. \qquad
\end{eqnarray}
The integration over $x$ is trivial:
\begin{equation}
\int_0^1 dy \int_{1-y}^1 dx \,
\frac{\partial}{\partial x} \frac{\partial}{\partial y}
{\cal F}_K \left( x, y \right)
= \int_0^1 dy \, \frac{\partial}{\partial y}
{\cal F}_K \left( 1, y \right)
- \int_0^1 dy \, \frac{\partial}{\partial y}
{\cal F}_K \left( x, y \right) \biggr|_{x=1-y} \,.
\end{equation}
The first term is
\begin{equation}
\int_0^1 dy \, \frac{\partial}{\partial y}
{\cal F}_K \left( 1, y \right)
= {\cal F}_K \left( 1, 1 \right) - {\cal F}_K \left( 1, 0 \right)
= {\cal F}_K (1,1) = \frac{2}{\pi^2} + \frac{1}{2} \,.
\end{equation}
The second term can also be calculated directly:
\begin{equation}
{}- \int_0^1 dy \, \frac{\partial}{\partial y}
{\cal F}_K \left( x, y \right) \biggr|_{x=1-y}
= {}- 4 \pi \int_0^1 dy \,\,
(1-y)^2 y \, \sin 2 \, \pi y = - \frac{3}{\pi^2} \,.
\end{equation}
It can also be calculated in the following way.
Let us first write
\begin{equation}
- \int_0^1 dy \, \frac{\partial}{\partial y}
{\cal F}_K \left( x, y \right) \biggr|_{x=1-y}
= - \int_0^1 dt \, \partial_y
{\cal F}_K \left( x, y \right) \biggr|_{x=1-t ,\, y=t} \,.
\end{equation}
Using the relation
\begin{equation}
{}[ \, x \, \partial_x + y \, \partial_y \, ] \,
{\cal F}_K (x,y) = 0 \,,
\end{equation}
which follows from (\ref{F_K-identity}), and
\begin{equation}
\partial_t {\cal F}_K \left( 1-t, t \right)
= {}- \partial_x {\cal F}_K \left( x, y \right)
\biggr|_{x=1-t ,\, y=t}
+ \partial_y {\cal F}_K \left( x, y \right)
\biggr|_{x=1-t ,\, y=t} \,,
\end{equation}
we find
\begin{equation}
\partial_y {\cal F}_K (x,y)
\biggr|_{x=1-t, \, y=t}
= (1-t) \, \partial_t \, {\cal F}_K (1-t,t)
= {} [ \, \partial_t \, (1-t) + 1 \, ] \, {\cal F}_K (1-t,t) \,.
\end{equation}
Since
\begin{equation}
{}- \int_0^1 dt \, \partial_t
\Bigl[ \, (1-t) \, {\cal F}_K \left( 1-t, t \right) \, \Bigr]
= {\cal F}_K (1,0) = 0 \,,
\end{equation}
we obtain
\begin{equation}
- \int_0^1 dy \, \frac{\partial}{\partial y}
{\cal F}_K \left( x, y \right) \biggr|_{x=1-y}
= - \int_0^1 dt \, {\cal F}_K \left( 1-t, t \right) \,.
\end{equation}
It is straightforward to carry out the integral:
\begin{eqnarray}
&& {}- \int_0^1 dt \, {\cal F}_K \left( 1-t, t \right)
\nonumber \\
&=& - \int_0^1 dt \, \biggl[ \, \frac{1}{\pi^2} \,
\Bigl( 1 + \cos ( \pi -2 \, \pi t ) \Bigr)
+ 2 \, (1-t) \, t \, 
\cos ( \pi -2 \, \pi t )
+ \frac{1-2 \, t}{\pi} \sin ( \pi -2 \, \pi t ) \, \biggr]
= {}- \frac{3}{\pi^2} \,.
\nonumber \\
\end{eqnarray}
The quantity ${\cal K}_2$ is therefore given by
\begin{equation}
{\cal K}_2
= {\cal F}_K (1,1) - \frac{3}{\pi^2}
= \frac{1}{2} - \frac{1}{\pi^2} \,.
\end{equation}
The difference between ${\cal K}_2$
and the value (\ref{prediction-kinetic})
predicted from Sen's conjecture
is written in terms of ${\cal F}_K (1,1)$,
and it is easy to see the way it is canceled
in (\ref{prediction-kinetic-reproduced}) from
our method of the calculation.

\section{Cubic term}
\label{Cubic-term-section}
\setcounter{equation}{0}

In order to evaluate the cubic term
of the Witten's string field theory action
for Schnabl's solution (\ref{Psi}),
we need to calculate the following quantities:
\begin{eqnarray}
{\cal V}_3
&=& \lim_{N \to \infty} \sum_{n=0}^N \sum_{m=0}^N \sum_{k=0}^N \,
\langle \, \psi'_n , \psi'_m \ast \psi'_k \, \rangle \,,
\\
{\cal V}_2
&=& \lim_{N \to \infty} \sum_{m=0}^N \sum_{k=0}^N \,
\langle \, \psi_N , \psi'_m \ast \psi'_k \, \rangle \,,
\\
{\cal V}_1
&=& \lim_{N \to \infty} \sum_{n=0}^N \,
\langle \, \psi'_n , \psi_N \ast \psi_N \, \rangle \,,
\\
{\cal V}_0
&=& \lim_{N \to \infty}
\langle \, \psi_N , \psi_N \ast \psi_N \, \rangle \,,
\phantom{\sum_{n=0}^N}
\end{eqnarray}
where the subscript of ${\cal V}$ again indicates
the number of sums.
These quantities all consist of
inner products of the form
$\langle \, \psi_n , \psi_m \ast \psi_k \, \rangle$
and their derivatives.
We calculate the inner product
$\langle \, \psi_n , \psi_m \ast \psi_k \, \rangle$
in the first subsection
and then calculate the summations
in the second subsection.

\subsection{Correlation functions}

\begin{figure}[htb]
\centerline{\epsfxsize=6in\epsfbox{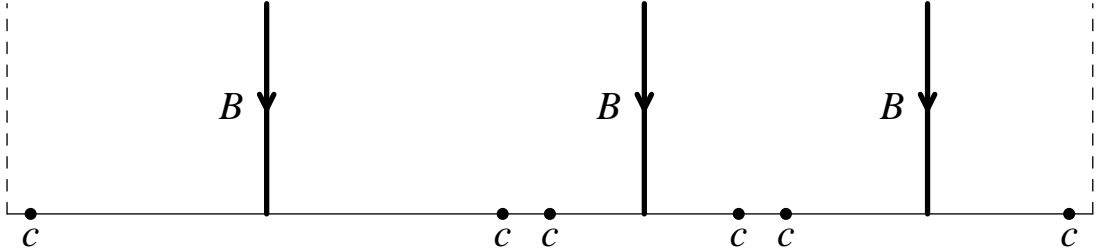}}
\caption{\small
A representation of the inner product
$\langle \, \psi_n , \psi_m \ast \psi_k \, \rangle$.
The two dashed lines are identified.
Each of the three states is represented
by figure \ref{figure2} with the shaded region deleted.
The semi-infinite cylinder in this figure
is constructed by gluing together the resulting three surfaces.
The $c$ ghost near the left dashed line has been brought
to the right in (\ref{psi_n-psi_m-psi_k}) using the periodicity.}
\label{figure4}
\end{figure}

The inner product
$\langle \, \psi_n , \psi_m \ast \psi_k \, \rangle$ is given by
\begin{eqnarray}
&& \langle \, \psi_n , \psi_m \ast \psi_k \, \rangle
\nonumber \\
&=& \left( \frac{2}{\pi} \right)^3 \,
\langle \, B \, c (x-a) \, c (x) \,
B \, c (x+y-a) \, c (x+y) \,
B \, c (x+y+z-a) \, c (x+y+z) \, \rangle_{C_{x+y+z}}
\label{psi_n-psi_m-psi_k}
\nonumber \\
\end{eqnarray}
with
\begin{equation}
x = \frac{\pi n}{2} + \frac{\pi}{2} \,, \qquad
y = \frac{\pi m}{2} + \frac{\pi}{2} \,, \qquad
z = \frac{\pi k}{2} + \frac{\pi}{2} \,, \qquad
a = \frac{\pi}{2} \,.
\end{equation}
See figure \ref{figure4}.
The calculation of this correlation function
can be reduced to that of
the three-point function of $c$ ghosts
on the same semi-infinite cylinder,
as we will demonstrate below.
The three-point function of $c$ ghosts
on the semi-infinite cylinder $C_\pi$
can be obtained from (\ref{normalization})
by the conformal transformation $z = f_\infty (w) = \arctan w$
and is given by
\begin{equation}
\langle \, c (z_1) \, c (z_2) \, c (z_3) \, \rangle_{C_\pi}
= \sin (z_1-z_2) \sin (z_1-z_3) \sin (z_2-z_3) \,.
\end{equation}
The three-point function of $c$ ghosts
on a general semi-infinite cylinder $C_n$ is
\begin{equation}
\langle \, c (z_1) \, c (z_2) \, c (z_3) \, \rangle_{C_n}
= \left( \frac{n}{\pi} \right)^3 \,
\sin \frac{\pi (z_1-z_2)}{n} \sin \frac{\pi (z_1-z_3)}{n}
\sin \frac{\pi (z_2-z_3)}{n} \,.
\end{equation}

Let us next calculate the correlation function
$\langle \, B \, c (z_1) \, c (z_2) \,
c (z_3) \, c (z_4) \, \rangle_{C_\pi}$.
One way of calculating this is
to carry out the integral of the $b$ ghost in $B$ explicitly,
for example, on the upper-half plane.
It can also be obtained in the following indirect way.
Since
\begin{equation}
\langle \, B \, c (z_1) \, c (z_2) \,
c (z_3) \, c (z_4) \, \rangle_{C_\pi}
= \frac{1}{2} \, \langle \, B \, c (z_1) \, c (z_2) \,
c (z_3) \, c (z_4) \, \rangle_{C_\pi}
+ \frac{1}{2} \, \langle \, c (z_1) \, c (z_2) \,
c (z_3) \, c (z_4) \, B \, \rangle_{C_\pi} \,,
\end{equation}
the integral of the $b$ ghost can be written
after the conformal transformation $\xi = \tan z$
to the upper-half plane with the coordinate $\xi$ as
\begin{equation}
\frac{1}{2} \left( -B_1^R+B_1^L \right)
= {}- \frac{1}{2} \, \oint \frac{d \xi}{2 \pi i} \,
(\xi^2+1) \, \varepsilon ( \, {\rm Re} \, \xi \, ) \, b (\xi) \,,
\end{equation}
where ${\rm Re} \, \xi$ is the real part of $\xi$,
and the step function $\varepsilon (x)$
is defined to be $\varepsilon (x) = 1$ for $x > 0$
and $\varepsilon (x) = -1$ for $x < 0$.
The contour of the integral should encircle
all of the four $c$ ghosts counterclockwise.
Furthermore,
\begin{equation}
\frac{1}{2} \left( -B_1^R+B_1^L \right)
= {}- \frac{1}{\pi} \, \oint \frac{d \xi}{2 \pi i} \,
(\xi^2+1) \, ( \arctan \xi + {\rm arccot} \, \xi ) \, b (\xi)
= {}- \frac{1}{\pi}
\left( {\cal B}_0 + {\cal B}_0^\dagger \right) \,,
\end{equation}
where
\begin{equation}
{\cal B}_0 = \oint \frac{d \xi}{2 \pi i} \,
(\xi^2+1) \, \arctan \xi \, b (\xi) \,,
\end{equation}
and ${\cal B}_0^\dagger$ is its BPZ conjugate.
There is no contribution from the ${\cal B}_0^\dagger$ part
because the integrand is regular at infinity
and there are no operator insertions outside the contour.
The integrand of ${\cal B}_0$ is regular at the origin,
but there are four $c$-ghost insertions inside the contour.
Since $c(z)$ is mapped to $\cos^2 z \, c(\tan z)$
in the $\xi$ coordinate, the contribution from each $c$-ghost
insertion is
\begin{equation}
\cos^2 z \left( - \frac{1}{\pi} \right)
\oint \frac{d \xi}{2 \pi i} \,
(\xi^2+1) \, \arctan \xi \, b (\xi) \, c(\tan z)
= -\frac{z}{\pi} \,.
\end{equation}
After mapping the upper-half plane back to $C_\pi$
and taking into account the signs from anticommuting ghosts,
the correlation function is given by
\begin{eqnarray}
\langle \, B \, c (z_1) \, c (z_2) \,
c (z_3) \, c (z_4) \, \rangle_{C_\pi}
&=& {}- \frac{z_1}{\pi} \,
\langle \, c (z_2) \,
c (z_3) \, c (z_4) \, \rangle_{C_\pi}
+ \frac{z_2}{\pi} \,
\langle \, c (z_1) \,
c (z_3) \, c (z_4) \, \rangle_{C_\pi}
\nonumber \\
&& {}- \frac{z_3}{\pi} \,
\langle \, c (z_1) \, c (z_2) \,
c (z_4) \, \rangle_{C_\pi}
+ \frac{z_4}{\pi} \,
\langle \, c (z_1) \, c (z_2) \,
c (z_3) \, \rangle_{C_\pi} \,. \qquad
\end{eqnarray}
The correlation function
$\langle \, B \, c (z_1) \, c (z_2) \,
c (z_3) \, c (z_4) \, \rangle_{C_n}$
on a general semi-infinite cylinder $C_n$ is
\begin{eqnarray}
\langle \, B \, c (z_1) \, c (z_2) \,
c (z_3) \, c (z_4) \, \rangle_{C_n}
&=& {}- \frac{z_1}{n} \,
\langle \, c (z_2) \,
c (z_3) \, c (z_4) \, \rangle_{C_n}
+ \frac{z_2}{n} \,
\langle \, c (z_1) \,
c (z_3) \, c (z_4) \, \rangle_{C_n}
\nonumber \\
&& {}- \frac{z_3}{n} \,
\langle \, c (z_1) \, c (z_2) \,
c (z_4) \, \rangle_{C_n}
+ \frac{z_4}{n} \,
\langle \, c (z_1) \, c (z_2) \,
c (z_3) \, \rangle_{C_n} \,.
\end{eqnarray}

Using the relation (\ref{B-c-c-B}),
the correlation function with three $B$ insertions
reduces to
\begin{eqnarray}
&& \langle \, B \, c (z_1) \, c (z_2) \,
B \, c (z_3) \, c (z_4) \,
B \, c (z_5) \, c (z_6) \, \rangle_{C_{\pi}}
\nonumber \\
&=& \langle \, B \, c (z_1) \, c (z_2) \,
B \, c (z_3) \, c (z_5) \, c (z_6) \, \rangle_{C_{\pi}}
- \langle \, B \, c (z_1) \, c (z_2) \,
B \, c (z_4) \, c (z_5) \, c (z_6) \, \rangle_{C_{\pi}}
\nonumber \\
&=& \langle \, B \, c (z_1) \,
c (z_3) \, c (z_5) \, c (z_6) \, \rangle_{C_{\pi}}
- \langle \, B \, c (z_2) \,
c (z_3) \, c (z_5) \, c (z_6) \, \rangle_{C_{\pi}}
\nonumber \\
&& {}- \langle \, B \, c (z_1) \,
c (z_4) \, c (z_5) \, c (z_6) \, \rangle_{C_{\pi}}
+ \langle \, B \, c (z_2) \,
c (z_4) \, c (z_5) \, c (z_6) \, \rangle_{C_{\pi}} \,.
\end{eqnarray}
The expression significantly simplifies
in the following case which is of our interest:
\begin{eqnarray}
&& \langle \, B \, c (z_1) \, c (z_1+a) \,
B \, c (z_2) \, c (z_2+a) \,
B \, c (z_3) \, c (z_3+a) \, \rangle_{C_{\pi}}
\nonumber \\
&=& {} \frac{4 \, a}{\pi} \, \sin^2 a \,
\sin (z_1-z_2) \, \sin (z_1-z_3) \, \sin (z_2-z_3) \,,
\end{eqnarray}
where we have used the formulas
\begin{eqnarray}
&& \sin (x+a) \, \sin (y+a) - \sin x \, \sin y
= \sin a \, \sin (x+y+a) \,,
\\
&& \sin (x+a) \, \sin y - \sin x \, \sin (y+a)
= - \sin a \, \sin (x-y) \,,
\\
&& \sin (2x-2y) + \sin (2y-2z) + \sin (2z-2x)
= 4 \, \sin (x-y) \, \sin (x-z) \, \sin (y-z) \,.
\end{eqnarray}
It is interesting to note that
\begin{equation}
\langle \, B \, c (z_1) \, c (z_1+a) \,
B \, c (z_2) \, c (z_2+a) \,
B \, c (z_3) \, c (z_3+a) \, \rangle_{C_{\pi}}
= \frac{4 \, a}{\pi} \, \sin^2 a \,
\langle \, c (z_1) \, c (z_2) \, c (z_3) \, \rangle_{C_{\pi}} \,.
\end{equation}
The correlation function on $C_n$ is given by
\begin{eqnarray}
&& \langle \, B \, c (z_1) \, c (z_1+a) \,
B \, c (z_2) \, c (z_2+a) \,
B \, c (z_3) \, c (z_3+a) \, \rangle_{C_n}
\nonumber \\
&=& \frac{4 \, a \, n^2}{\pi^3} \, \sin^2 \frac{\pi a}{n} \,
\sin \frac{\pi (z_1-z_2)}{n} \, \sin \frac{\pi (z_1-z_3)}{n} \,
\sin \frac{\pi (z_2-z_3)}{n} \,,
\end{eqnarray}
and the inner product
$\langle \, \psi_n , \psi_m \ast \psi_k \, \rangle$ is
\begin{eqnarray}
&& \langle \, \psi_n , \psi_m \ast \psi_k \, \rangle
\nonumber \\
&=& {}- \left( \frac{2}{\pi} \right)^3 \,
\frac{4 \, a \, (x+y+z)^2}{\pi^3} \,
\sin^2 \frac{\pi a}{x+y+z} \,
\sin \frac{\pi x}{x+y+z} \, \sin \frac{\pi y}{x+y+z} \,
\sin \frac{\pi z}{x+y+z}
\end{eqnarray}
with
\begin{equation}
x = \frac{\pi n}{2} + \frac{\pi}{2} \,, \qquad
y = \frac{\pi m}{2} + \frac{\pi}{2} \,, \qquad
z = \frac{\pi k}{2} + \frac{\pi}{2} \,, \qquad
a = \frac{\pi}{2} \,.
\end{equation}

\subsection{Summations}

In the proofs of (\ref{psi'_0-relation})
and (\ref{psi'_n-relation})
in section \ref{Equation-of-motion},
it was assumed that these equations are contracted with
a state in the Fock space.
However, it is straightforward to see
in the CFT formulation, for example, that
the proofs can be extended to the case where
the equations are contracted with $\psi'_k$
for any $k$ in the range $k \ge 0$:
\begin{equation}
\langle \, \psi'_k , Q_B \psi'_0 \, \rangle = 0 \,, \qquad
\langle \, \psi'_k , Q_B \psi'_{n+1} \, \rangle
= - \sum_{m=0}^{n} \langle \, \psi'_k ,
\psi'_{m} \ast \psi'_{n-m} \, \rangle \,.
\end{equation}
Combining these with (\ref{psi'-Q_B-psi'}), we obtain
\begin{equation}
\sum_{m=0}^n \sum_{k=0}^{n-m}
\langle \, \psi'_m , \psi'_k \ast \psi'_{n-m-k} \, \rangle = 0
\label{psi'-psi'-psi'}
\end{equation}
for any nonnegative integer $n$.
The triple sum in ${\cal V}_3$
before taking the limit $N \to \infty$
can be decomposed in the following way:
\begin{eqnarray}
&& \sum_{n=0}^N \sum_{m=0}^N \sum_{k=0}^N \,
\langle \, \psi'_n , \psi'_m \ast \psi'_k \, \rangle
\nonumber \\
&=& \sum_{k=0}^N \sum_{m=0}^{N-k} \sum_{n=0}^{N-k-m} \,
\langle \, \psi'_n , \psi'_m \ast \psi'_k \, \rangle
\nonumber \\
&& {}+ \sum_{k=0}^N \sum_{m=0}^{N-k} \sum_{n=N-k-m+1}^N \,
\langle \, \psi'_n , \psi'_m \ast \psi'_k \, \rangle
+ \sum_{k=1}^N \sum_{m=N-k+1}^N \sum_{n=0}^N \,
\langle \, \psi'_n , \psi'_m \ast \psi'_k \, \rangle
\nonumber \\
&=& \sum_{n=0}^N \sum_{m=0}^n \sum_{k=0}^{n-m} \,
\langle \, \psi'_m , \psi'_k \ast \psi'_{n-m-k} \, \rangle
\nonumber \\
&& {}+ \sum_{k=0}^N \sum_{m=0}^{N-k} \sum_{n=N-k-m+1}^N \,
\langle \, \psi'_n , \psi'_m \ast \psi'_k \, \rangle
+ \sum_{k=1}^N \sum_{m=N-k+1}^N \sum_{n=0}^N \,
\langle \, \psi'_n , \psi'_m \ast \psi'_k \, \rangle
\label{V_3-decomposition}
\end{eqnarray}
with the understanding that there is no contribution
to the second triple sum on the right-hand side when $k=m=0$.
The first triple sum
on the right-hand side of (\ref{V_3-decomposition})
vanishes because of (\ref{psi'-psi'-psi'}).
The quantity ${\cal V}_3$ is thus given by
\begin{equation}
{\cal V}_3 = \lim_{N \to \infty}
\sum_{k=0}^N \sum_{m=0}^{N-k} \sum_{n=N-k-m+1}^N \,
\langle \, \psi'_n , \psi'_m \ast \psi'_k \, \rangle
+ \lim_{N \to \infty}
\sum_{k=1}^N \sum_{m=N-k+1}^N \sum_{n=0}^N \,
\langle \, \psi'_n , \psi'_m \ast \psi'_k \, \rangle \,.
\label{reduced-V_3}
\end{equation}
As in the case of the evaluation of the kinetic term
in the previous section,
all of ${\cal V}_3$, ${\cal V}_2$, ${\cal V}_1$, and ${\cal V}_0$
are now written in terms of inner products of the form
$\langle \, \psi_n , \psi_m \ast \psi_k \, \rangle$
with $n+m+k$ large and their derivatives.
When $n+m+k$ is large, the inner product
$\langle \, \psi_n , \psi_m \ast \psi_k \, \rangle$ is given by
\begin{equation}
\lim_{n+m+k \to \infty} \,
\langle \, \psi_n , \psi_m \ast \psi_k \, \rangle
= {}- \frac{4}{\pi} \,
\sin \frac{\pi n}{n+m+k} \, \sin \frac{\pi m}{n+m+k} \,
\sin \frac{\pi k}{n+m+k} \,.
\end{equation}
Let us introduce the function
\begin{equation}
{\cal F}_V (x,y,z) = {}- \frac{4}{\pi} \,
\sin \frac{\pi x}{x+y+z} \, \sin \frac{\pi y}{x+y+z} \,
\sin \frac{\pi z}{x+y+z} \,.
\end{equation}
Note that
\begin{equation}
{\cal F}_V (0,y,z) = {\cal F}_V (x,0,z)
= {\cal F}_V (x,y,0) = 0
\end{equation}
and
\begin{equation}
{\cal F}_V (ax, ay, az) = {\cal F}_V (x,y,z)
\label{F_V-identity}
\end{equation}
for any nonvanishing $a$.
Then the inner product
$\langle \, \psi_n , \psi_m \ast \psi_k \, \rangle$
in the limit $n+m+k \to \infty$ can be written as
\begin{equation}
\lim_{n+m+k \to \infty}
\langle \, \psi_n , \psi_m \ast \psi_k \, \rangle
= {\cal F}_V \left( a n, a m, a k \right)
\end{equation}
for any nonvanishing $a$.
We can in particular choose $a$ to be $1/N$
in taking the limit $N \to \infty$:
\begin{equation}
\lim_{n+m+k \to \infty}
\langle \, \psi_n , \psi_m \ast \psi_k \, \rangle
= {\cal F}_V
\left( \frac{n}{N}, \frac{m}{N}, \frac{k}{N} \right) \,.
\end{equation}

The quantity ${\cal V}_0$ is readily given by
\begin{equation}
{\cal V}_0 = \lim_{N \to \infty}
\langle \, \psi_N , \psi_N \ast \psi_N \, \rangle
= {\cal F}_V (1,1,1)
= {}- \frac{3 \sqrt{3}}{2 \, \pi} \,.
\end{equation}
The sums in ${\cal V}_1$ and ${\cal V}_2$
can be transformed into integrals.
The quantity ${\cal V}_1$ is given by
\begin{eqnarray}
{\cal V}_1
&=& \lim_{N \to \infty} \sum_{n=0}^{N} \frac{\partial}{\partial n}
\langle \, \psi_n , \psi_N \ast \psi_N \, \rangle
= \lim_{N \to \infty} \sum_{n=0}^{N} \frac{\partial}{\partial n}
{\cal F}_V \left( \frac{n}{N}, 1, 1 \right)
\nonumber \\
&=& \lim_{N \to \infty} \frac{1}{N} \sum_{n=0}^{N}
\frac{\partial}{\partial x} {\cal F}_V \left( x, 1, 1 \right)
\biggr|_{x=\frac{n}{N}}
= \int_0^1 dx \,
\frac{\partial}{\partial x} {\cal F}_V \left( x, 1, 1 \right)
= {\cal F}_V (1,1,1) = {}- \frac{3 \sqrt{3}}{2 \, \pi} \,, \qquad
\end{eqnarray}
and ${\cal V}_2$ is
\begin{eqnarray}
{\cal V}_2
&=& \lim_{N \to \infty} \sum_{m=0}^{N} \sum_{k=0}^{N}
\frac{\partial}{\partial m} \frac{\partial}{\partial k}
\langle \, \psi_N , \psi_m \ast \psi_k \, \rangle
= \lim_{N \to \infty} \sum_{m=0}^{N} \sum_{k=0}^{N}
\frac{\partial}{\partial m} \frac{\partial}{\partial k}
{\cal F}_V \left( 1, \frac{m}{N}, \frac{k}{N} \right)
\nonumber \\
&=& \lim_{N \to \infty} \frac{1}{N^2}
\sum_{m=0}^{N} \sum_{k=0}^{N}
\frac{\partial}{\partial y} \frac{\partial}{\partial z}
{\cal F}_V \left( 1, y, z \right)
\biggr|_{y=\frac{m}{N}, \, z=\frac{k}{N}}
\nonumber \\
&=& \int_0^1 dy \int_0^1 dz \,
\frac{\partial}{\partial y} \frac{\partial}{\partial z}
{\cal F}_V \left( 1, y, z \right)
= \int_0^1 dy \, \frac{\partial}{\partial y}
{\cal F}_V \left( 1, y, 1 \right)
= {\cal F}_V (1,1,1)
= {}- \frac{3 \sqrt{3}}{2 \, \pi} \,.
\end{eqnarray}
Finally, ${\cal V}_3$ in the form of (\ref{reduced-V_3})
can be expressed as a sum of two integrals:
\begin{equation}	
{\cal V}_3
= \int_0^1 dz \int_0^{1-z} dy \int_{1-z-y}^1 dx \,
\frac{\partial}{\partial x} \frac{\partial}{\partial y}
\frac{\partial}{\partial z} {\cal F}_V \left( x, y, z \right)
+ \int_0^1 dz \int_{1-z}^1 dy \int_0^1 dx \,
\frac{\partial}{\partial x} \frac{\partial}{\partial y}
\frac{\partial}{\partial z} {\cal F}_V \left( x, y, z \right) \,.
\end{equation}
The integration over $x$ is trivial:
\begin{eqnarray}
{\cal V}_3
&=& \int_0^1 dz \int_0^{1-z} dy \,
\frac{\partial}{\partial y} \frac{\partial}{\partial z}
{\cal F}_V \left( 1, y, z \right)
- \int_0^1 dz \int_0^{1-z} dy \,
\frac{\partial}{\partial y} \frac{\partial}{\partial z}
{\cal F}_V \left( x, y, z \right) \biggr|_{x=1-z-y}
\nonumber \\
&& {}+ \int_0^1 dz \int_{1-z}^1 dy \,
\frac{\partial}{\partial y} \frac{\partial}{\partial z}
{\cal F}_V \left( 1, y, z \right)
\nonumber \\
&=& \int_0^1 dz \int_0^1 dy \,
\frac{\partial}{\partial y} \frac{\partial}{\partial z}
{\cal F}_V \left( 1, y, z \right)
- \int_0^1 dz \int_0^{1-z} dy \,
\frac{\partial}{\partial y} \frac{\partial}{\partial z}
{\cal F}_V \left( x, y, z \right) \biggr|_{x=1-z-y} \,.
\end{eqnarray}
The first term is
\begin{equation}
\int_0^1 dz \int_0^1 dy \,
\frac{\partial}{\partial y} \frac{\partial}{\partial z}
{\cal F}_V \left( 1, y, z \right) = {\cal F}_V (1,1,1)
= {}- \frac{3 \sqrt{3}}{2 \, \pi} \,.
\end{equation}
While it is also possible to calculate the second term directly,
we transform it in the following way:
\begin{equation}
- \int_0^1 dz \int_0^{1-z} dy \,
\frac{\partial}{\partial y} \frac{\partial}{\partial z}
{\cal F}_V \left( x, y, z \right) \biggr|_{x=1-z-y}
= - 2 \int_0^1 dv \int_0^{1-v} du \,
{\cal F}_V \left( 1-v-u, u, v \right) \,.
\label{integral}
\end{equation}
A derivation of (\ref{integral}) is given
in appendix \ref{appendix}.
Since
\begin{eqnarray}
{\cal F}_V ( 1-v-u, u, v) &=& {}- \frac{4}{\pi} \,
\sin \left[ \, \pi (1-v-u) \, \right] \, \sin \pi u \, \sin \pi v
\nonumber \\
&=& {}- \frac{1}{\pi} \Bigl[ \, \sin
\left[ \, 2 \pi (1-v-u) \, \right]
+ \sin 2 \pi u  + \sin 2 \pi v \, \Bigr]
\nonumber \\
&=& \frac{1}{\pi} \Bigl[ \, \sin \left[ \, 2 \pi (u+v) \, \right]
- \sin 2 \pi u  - \sin 2 \pi v \, \Bigr] \,,
\end{eqnarray}
it is straightforward to calculate the integral:
\begin{eqnarray}
&& {} - 2 \int_0^1 dv \int_0^{1-v} du \,
{\cal F}_V \left( 1-v-u, u, v \right)
\nonumber \\
&=& {}- \frac{2}{\pi} \int_0^1 dv \int_0^{1-v} du \,
\Bigl[ \, \sin \left[ \, 2 \pi (u+v) \, \right]
- \sin 2 \pi u  - \sin 2 \pi v \, \Bigr]
= \frac{3}{\pi^2} \,.
\end{eqnarray}
The quantity ${\cal V}_3$ is therefore given by
\begin{equation}
{\cal V}_3 = {\cal F}_V (1,1,1) + \frac{3}{\pi^2}
= \frac{3}{\pi^2} - \frac{3 \sqrt{3}}{2 \, \pi} \,.
\end{equation}

\section{Conclusions}
\label{Conclusions}
\setcounter{equation}{0}

We found that
the equation of motion contracted with the solution itself
(\ref{equation-with-Psi}) is not satisfied
without the $\psi_N$ piece in (\ref{Psi}) because
\begin{eqnarray}
{\cal K}_2
&=& \lim_{N \to \infty} \sum_{n=0}^N \sum_{m=0}^N \,
\langle \, \psi'_n , Q_B \psi'_m \, \rangle
= {}- \frac{3}{\pi^2} + {\cal F}_K (1,1) \,,
\\
{\cal V}_3
&=& \lim_{N \to \infty} \sum_{n=0}^N \sum_{m=0}^N \sum_{k=0}^N \,
\langle \, \psi'_n , \psi'_m \ast \psi'_k \, \rangle
= \frac{3}{\pi^2} + {\cal F}_V (1,1,1) \,,
\end{eqnarray}
and ${\cal F}_K (1,1) + {\cal F}_V (1,1,1) \ne 0$,
but it is satisfied
when the $\psi_N$ piece is included.
The term ${\cal F}_K (1,1)$ in ${\cal K}_2$
and the term ${\cal F}_V (1,1,1)$ in ${\cal V}_3$
are canceled when the $\psi_N$ piece is added,
and the values (\ref{prediction-kinetic})
and (\ref{prediction-cubic}) predicted by Sen's conjecture
are nontrivially reproduced by the following integrals:
\begin{eqnarray}
\langle \, \Psi , Q_B \Psi \, \rangle
&=& {}- \int_0^1 dt \, {\cal F}_K \left( 1-t, t \right)
= {}- \frac{3}{\pi^2} \,,
\label{kinetic-integral}
\\
\langle \, \Psi , \Psi \ast \Psi \, \rangle \,
&=& - 2 \int_0^1 dv \int_0^{1-v} du \,
{\cal F}_V \left( 1-v-u, u, v \right)
= \frac{3}{\pi^2} \,.
\label{cubic-integral}
\end{eqnarray}
We emphasize that the cancellation
between (\ref{kinetic-integral})
and (\ref{cubic-integral}) is not a consequence
of (\ref{equation-with-phi}).
We in fact do not have any deep understanding
of why these apparently unrelated integrals should cancel,
but it is really necessary for Schnabl's solution (\ref{Psi})
to work out.
It would be interesting, for example, if we could understand
Schnabl's solution better
in the context of noncommutative geometry underlying
Witten's open cubic string field theory \cite{Witten:1985cc}.

\section*{Acknowledgments}
I would like to thank Ian Ellwood, Yoichi Kazama, Martin Schnabl,
Wati Taylor, and Barton Zwiebach for useful discussions.
I would also like to express my gratitude
to Barton Zwiebach for his helpful comments and suggestions
on an earlier version of the manuscript.
This work is supported in part by funds
provided by the U.S. Department of Energy (D.O.E.)
under cooperative research agreement DE-FC02-94ER40818.

\appendix
\renewcommand{\theequation}{\Alph{section}.\arabic{equation}}

\section{Integral}
\label{appendix}
\setcounter{equation}{0}

We derive (\ref{integral}) in this appendix.
Let us first write
\begin{equation}
- \int_0^1 dz \int_0^{1-z} dy \,
\frac{\partial}{\partial y} \frac{\partial}{\partial z}
{\cal F}_V \left( x, y, z \right) \biggr|_{x=1-z-y}
= {}- \int_0^1 dv \int_0^{1-v} du \, \partial_y \partial_z
{\cal F}_V \left( x, y, z \right)
\biggr|_{x=1-v-u, \, y=u, \, z=v} \,.
\end{equation}
Using the relation
\begin{equation}
{}[ \, x \, \partial_x + y \, \partial_y + z \, \partial_z
+ 1 \, ] \, \partial_z {\cal F}_V (x,y,z) = 0 \,,
\end{equation}
which follows from (\ref{F_V-identity}), and
\begin{eqnarray}
\partial_u \partial_z
{\cal F}_V \left( x, y, z \right) \biggr|_{x=1-v-u, \, y=u, \, z=v}
&=& ( - \partial_x + \partial_y ) \, \partial_z
{\cal F}_V \left( x, y, z \right)
\biggr|_{x=1-v-u, \, y=u, \, z=v} \,,
\\
\partial_v \partial_z
{\cal F}_V \left( x, y, z \right) \biggr|_{x=1-v-u, \, y=u, \, z=v}
&=& (- \partial_x + \partial_z ) \, \partial_z
{\cal F}_V \left( x, y, z \right)
\biggr|_{x=1-v-u, \, y=u, \, z=v} \,,
\end{eqnarray}
we obtain
\begin{eqnarray}
\partial_y \partial_z {\cal F}_V (x,y,z)
\biggr|_{x=1-v-u, \, y=u, \, z=v}
&=& {} [ \, (1-u) \, \partial_u - v \, \partial_v - 1 \, ] \,
\partial_z {\cal F}_V (x,y,z)
\biggr|_{x=1-v-u, \, y=u, \, z=v}
\nonumber \\
&=& {} [ \, \partial_u \, (1-u) - \partial_v \, v + 1 \, ] \,
\partial_z {\cal F}_V (x,y,z)
\biggr|_{x=1-v-u, \, y=u, \, z=v} \,. \qquad
\end{eqnarray}
Note that
\begin{equation}
{}- \int_0^1 dv \int_0^{1-v} du \, \partial_u
\left[ \, (1-u) \, \partial_z {\cal F}_V \left( x, y, z \right)
\biggr|_{x=1-v-u, \, y=u, \, z=v} \, \right] = 0 \,,
\end{equation}
and
\begin{eqnarray}
&& \int_0^1 dv \int_0^{1-v} du \, \partial_v
\left[ \, v \, \partial_z {\cal F}_V \left( x, y, z \right)
\biggr|_{x=1-v-u, \, y=u, \, z=v} \, \right]
\nonumber \\
&=& \int_0^1 du \int_0^{1-u} dv \, \partial_v
\left[ \, v \, \partial_z {\cal F}_V \left( x, y, z \right)
\biggr|_{x=1-v-u, \, y=u, \, z=v} \, \right] =0 \,,
\end{eqnarray}
where we used that
$\partial_z {\cal F}_V \left( 0, y, z \right) = 0$
and $\partial_z {\cal F}_V \left( x, 0, z \right) = 0$.
We therefore obtain
\begin{equation}
- \int_0^1 dv \int_0^{1-v} du \, \partial_y \partial_z
{\cal F}_V \left( x, y, z \right) \biggr|_{x=1-v-u, \, y=u, \, z=v}
= - \int_0^1 dv \int_0^{1-v} du \, \partial_z
{\cal F}_V \left( x, y, z \right)
\biggr|_{x=1-v-u, \, y=u, \, z=v} \,.
\end{equation}
Similarly, using the relation
\begin{equation}
{}[ \, x \, \partial_x + y \, \partial_y + z \, \partial_z \, ] \,
{\cal F}_V (x,y,z) = 0 \,,
\end{equation}
which follows from (\ref{F_V-identity}), and
\begin{eqnarray}
\partial_u
{\cal F}_V \left( 1-v-u, u, v \right)
&=& ( - \partial_x + \partial_y ) \,
{\cal F}_V \left( x, y, z \right)
\biggr|_{x=1-v-u, \, y=u, \, z=v} \,,
\\
\partial_v
{\cal F}_V \left( 1-v-u, u, v \right)
&=& (- \partial_x + \partial_z ) \,
{\cal F}_V \left( x, y, z \right)
\biggr|_{x=1-v-u, \, y=u, \, z=v} \,,
\end{eqnarray}
we obtain
\begin{eqnarray}
\partial_z {\cal F}_V (x,y,z)
\biggr|_{x=1-v-u, \, y=u, \, z=v}
&=& {} [ \, - u \, \partial_u + (1-v) \, \partial_v \, ] \,
{\cal F}_V (1-v-u,u,v)
\nonumber \\
&=& {} [ \, {}- \partial_u \, u + \partial_v \, (1-v) + 2 \, ] \,
{\cal F}_V (1-v-u,u,v) \,.
\end{eqnarray}
Note that
\begin{equation}
\int_0^1 dv \int_0^{1-v} du \, \partial_u
\Bigl[ \, u \, {\cal F}_V \left( 1-v-u, u, v \right) \, \Bigr]
= 0 \,,
\end{equation}
and
\begin{eqnarray}
&& {}- \int_0^1 dv \int_0^{1-v} du \, \partial_v
\Bigl[ \, (1-v) \, {\cal F}_V \left( 1-v-u, u, v \right) \,
\Bigr]
\nonumber \\
&=& {}- \int_0^1 du \int_0^{1-u} dv \, \partial_v
\Bigl[ \, (1-v) \, {\cal F}_V \left( 1-v-u, u, v \right) \,
\Bigr] =0 \,,
\end{eqnarray}
where we used that
${\cal F}_V \left( 0, y, z \right) = 0$
and 
${\cal F}_V \left( x, y, 0 \right) = 0$.
We have therefore derived that
\begin{eqnarray}
&& {}- \int_0^1 dv \int_0^{1-v} du \, \partial_y \partial_z
{\cal F}_V \left( x, y, z \right) \biggr|_{x=1-v-u, \, y=u, \, z=v}
\nonumber \\
&=& - \int_0^1 dv \int_0^{1-v} du \, \partial_z
{\cal F}_V \left( x, y, z \right) \biggr|_{x=1-v-u, \, y=u, \, z=v}
\nonumber \\
&=& - 2 \int_0^1 dv \int_0^{1-v} du \,
{\cal F}_V \left( 1-v-u, u, v \right) \,.
\end{eqnarray}


\end{document}